\documentclass{iopart}
\usepackage{epsf,amssymb,iopams,setstack}
\usepackage{graphicx}
\catcode`\@=11
\usepackage{epsfig}
\usepackage{epstopdf}

\newcommand \vev [1] {\langle{#1}\rangle}

\def\II{\hbox{{1}\kern-.25em\hbox{l}}}

\begin{document}
\title[Iterative construction of eigenfunctions]{Iterative construction of eigenfunctions of
the monodromy matrix for  $SL(2,\mathbb{C})$ magnet.}

\author{S. {\'E}. Derkachov$^{1,2}$ and A. N. Manashov$^{3,4}$
}

\address{$^1$ St.Petersburg Department of Steklov
Mathematical Institute of Russian Academy of Sciences,\\
\ \ \
Fontanka 27, 191023 St.Petersburg, Russia.
}
\address{$^2$ Dept. Appl. Math., St.~Petersburg State Polytechnic University,
Polytekhnicheskaya st.29, \\
\ \ \ 195251, St.Petersburg. }

\address{$^3$ Institute for Theoretical Physics, University of  Regensburg,
D-93040 Regensburg, Germany}
\address{$^4$ Department of Theoretical Physics,  Saint-Petersburg State University, \mbox{St.-Petersburg,} Russia}

\ead{derkach@pdmi.ras.ru}
\ead{alexander.manashov@physik.uni-regensburg.de}
\begin{abstract}
Eigenfunctions of the matrix elements of the monodromy matrix provide a convenient basis for studies of spin
chain models.
We present an iterative method for constructing  the eigenfunctions
in  the case of the
$SL(2,\mathbb{C})$ spin chains.
We derived an explicit integral representation for the eigenfunctions and calculated the corresponding scalar
products (Sklyanin's measure).
\end{abstract}

\maketitle

\setcounter{footnote}{0}

\section{Introduction}
The quantum inverse scattering method is a powerful tool for constructing and solving
integrable models. The fundamental object in this approach is the so-called
$\mathcal{R}-$matrix -- a linear operator which depends on a complex parameter (spectral parameter) and
satisfies a certain nonlinear relation known  as the Yang~-~Baxter equation (YBE). Each solution of this equation
gives rise to a family of  commuting operators. In many cases a commutative family includes an operator which can be
identified with a Hamiltonian of some physical system. The most famous example of an such integrable system is the
$\mathrm{XXX}_{1/2}$-spin chain -- the celebrated Heisenberg spin $1/2$ magnet solved by  H.~Bethe in
1931~\cite{Bethe:1931hc}. The general algebraic framework was developed much later and became known as Quantum
Inverse Scattering Method (QISM). For a review and references see Refs.~\cite{Baxter,Fad,FST,TaFa,KulSk,Skl91}.

Integrable  models with a finite dimensional Hilbert space such as spin  magnets of different types, found
many applications in  statistical and solid state physics~\cite{Baxter}. Quite unexpectedly spin magnets
arise also in the studies of high-energy scattering amplitudes in quantum field theories, namely in the
gauge field theories. Most of them can be solved with the help of the Algebraic Bethe
Ansatz(ABA)~\cite{Fad,FST,TaFa,KulSk,Skl91}. In this approach eigenstates of the model are constructed as
 excitations of certain type over the special (preudovacuum) state belonging to  the Hilbert space of the
system. However, there are integrable models, e.g. the Toda
chain~\cite{Gutzwiller:1981by,Sklyanin:1984sb,Gaudin,Kharchev:1999bh} and  the quantum KdV
model~\cite{Bazhanov:1994ft,Bazhanov:1996dr}, which can not be solved within the ABA. Such models have an
infinite~-~dimensional Hilbert space and the pseudovacuum state does not belong to it. Nevertheless they
can be solved by the methods of Baxter $\mathcal{Q}-$operators~\cite{Baxter:1972hz} and Separation of
Variables (SoV)~\cite{Sklyanin:1995bm}.

In the present work we consider another model of this type -- the so-called noncompact $SL(2,\mathbb{C})$ spin magnet.
Interest to such models stems from the  studies of Regge behaviour of  hadron scattering amplitudes,  for a review see
Ref.~\cite{Lipatov:1996ts}. It turns out that the Hamiltonian which governs the scale dependence of the scattering
amplitudes
in high~-~energy limit is integrable and can be identified with the Hamiltonian of a spin
magnet~\cite{Lipatov:1993qn,Lipatov:1993yb,Faddeev:1994zg}. This model was solved in
Refs.~\cite{DeVega:2001pu,deVega:2002im,Derkachov:2001yn,Derkachov:2002wz}
 with the help of Baxter $\mathcal{Q}-$operators and
SoV methods. Recently it was argued that the behaviour of  scattering amplitudes
in the multi-Regge kinematics in $\mathcal{N}=4$ SUSY
is governed by the Hamiltonian of the noncompact open spin chain~\cite{Lipatov:2009nt,Bartels:2011nz}.
The Hamiltonian of the model commutes with the diagonal entry of the monodromy matrix, $D(u)$.
In both cases, in order to diagonalize the Hamiltonian one has first to construct  eigenfunctions for  entries of the monodromy
matrix ($B$ or $D$). Let us also mention  that
the   problem of diagonalization of the operator $D$ for  finite dimensional representations of the $SL(2)$ group
 was addressed in Refs.\ \cite{MS,Terras}.

In this work we provide a regular recurrence procedure for constructing eigenfunctions for all entries of the monodromy
matrix. Our approach relies heavily on the representation  of the
$sl(2)-$invariant $\mathcal{R}-$matrix in the factorized form~\cite{Derkachov:2005hw,Derkachov:2010zz}.
The operators which factorize  the $\mathcal{R}-$matrix
play
a prominent role in our construction. Using them one can
 construct operators that intertwine the entries of the
monodromy matrix for the chains of  different length ($B_N(u)\Lambda_N\sim\Lambda_N B_{N-1}(u)$, an so on).
It immediately leads  to a  recurrence  construction.
We derive an integral  representation for the eigenfunctions and calculate their  scalar products (Sklyanin's measure).

It  was shown by Sklyanin~\cite{Skl91} that the eigenvalue equations for the transfer matrix for  the rank one chain models
become separated
in the basis provided by the eigenfunctions of the operator~$B_N(u)$.
At present time the SoV representation
is known for a variety of models. Among them are the Toda chain~\cite{Kharchev:1999bh,Kharchev:2000yj,Silantyev,Kozlowski},
different types of
$XXX$~\cite{Derkachov:2001yn,Derkachov:2002tf,Derkachov:2003qb,Niccoli:2012vq}
and   $XXZ$ spin
chains~\cite{BT,Niccoli:2011nj,Niccoli:2012ci,Faldella:2013qha}.

The paper is organized as follows: In Sect.~\ref{sect:preliminaries}
we describe the model and some basic elements of the QISM method. In Sect.\ref{sect:iterative} we develop  an
iterative procedure for constructing the eigenfunctions of the elements of the monodromy matrix.
In Sect.~\ref{sklyanin-measure} we calculate   scalar products of the eigenfunctions and
determine the Sklyanin measure. The method of constructing the Baxter operators is described in Sect.~\ref{sect:Baxter}.
The Hamiltonians for $D-$system are discussed in  Sect.~\ref{sect:Hamiltonian}.
Concluding remarks are presented in Sect.~\ref{sect:summary}. Several Appendices contain technical details.

\section{Preliminaries}\label{sect:preliminaries}

The quantum $SL(2,\mathbb{C})$ spin magnet is a straightforward generalization of the
standard $\mathrm{XXX}_{s}$ spin chain. In both models the dynamical variables are the spin operators,
$\vec{S}_k$, $k=1,\ldots,N$, where $N$ is the length of the chain.
In the $\mathrm{XXX}_{s}$ model
 the spin operators belong to a finite  dimensional representation
of the $SU(2)$ group so that the Hilbert space of the model is finite dimensional.
In the case of the $SL(2,\mathbb{C})$ spin magnet the spin generators belong to a
unitary continuous principal series representation of the $SL(2,\mathbb{C})$ group and the corresponding Hilbert
space is infinite dimensional.

The unitary principal series representation of the $SL(2,\mathbb{C})$ group, $T^{(s,\bar s)}$, is
determined by two complex  numbers (spins),
 $s$ and $\bar s$, such that $s-\bar s$ is a half-integer and $s+\bar s^*=1$~\cite{Gelfand}.
It acts on the space $L_2(\mathbb{C})$  and the group transformations take the form
\begin{equation}\label{Tg}
[T^{(s,\bar s)}(g^{-1})f](z,\bar z)=(cz+d)^{-2s}(\bar c \bar z+\bar d)^{-2\bar s}\,f\left(\frac{az+b}{cz+d},
\frac{\bar a\bar z+\bar b}{\bar c\bar z+\bar d}\right)\,.
\end{equation}
Here $g$ is a complex unimodular matrix,
$g=
\left(
\begin{array}{cc}
     a&b\\
c& d
\end{array}
\right)
$,
$ab-cd=1$, and $f\in L_2(\mathbb{C})$. For the unitary representations the spins $s$, $\bar s$
 can be parameterized as follows
\begin{eqnarray}
s=\frac{1+n_s}2+i\nu_s, \qquad\qquad \bar s=\frac{1-n_s}2+i\nu_s,
\end{eqnarray}
where $n_s$ is half-integer and $\nu_s$ is real.
The operators~(\ref{Tg})
are unitary with respect to the standard scalar product
\begin{equation}\label{sc}
(f,\psi)=\int d^2z \bar f(z)\psi(z),\qquad\qquad (T^{(s,\bar s)}(g)f,T^{(s,\bar s)}(g)\psi)=(f,\psi).
\end{equation}
The generators of infinitesimal transformations (spin operators) take the form
\begin{eqnarray}
S_-=-\partial_z, \qquad S_0=z\partial_z+s, \qquad S_+=z^2\partial_z+2s z\,,
\nonumber\\
\bar S_-=-\partial_{\bar z}, \qquad \bar S_0={\bar z}\partial_{\bar z}+\bar s, \qquad \bar S_+=\bar z^2\partial_{\bar z}+2\bar s \bar z\,
\end{eqnarray}
and satisfy the standard $sl(2)$ commutation relations
\begin{eqnarray}
[S_+,S_-]=2S_0, \qquad [S_0,S_\pm]=\pm S_\pm,\nonumber\\{}
 [\bar S_+,\bar S_-]=2\bar S_0, \qquad [\bar S_0,\bar S_\pm]=\pm \bar S_\pm.
\end{eqnarray}
The holomorphic ($S_\alpha$) and anti-holomorphic  ($\bar S_\alpha$) generators commute.
For  the unitary representations the holomorphic  and anti-holomorphic  generators are
adjoint to each other, $S_\alpha^\dagger=-\bar S_\alpha$.

Summarising: The quantum $SL(2,\mathbb{C})$ spin magnet
is  a one-dimensional lattice model.
The Hilbert space of the model is given by the direct product of the $L_2(\mathbb{C})$ spaces,
\begin{equation}\label{HN}
\mathbb{H}_N=\mathbb{V}_1\otimes \mathbb{V}_2\otimes\ldots \otimes \mathbb{V}_N,  \qquad \mathbb{V}_k=L_2(\mathbb{C})\,,\qquad
k=1,\ldots,N.
\end{equation}

The dynamical variables are given by two sets of spin operators~\footnote{It is assumed that
 the generators with index $k$ act non-trivially only on $k-$th space
in the tensor product, $\mathbb{V}_k$.}~--~holomorphic ( $S^{(k)}_{\pm,0}$ ) and anti-holomorphic ( $\bar
S^{(k)}_{\pm,0}$ ), $k=1,\ldots,N$.
In what follows we will consider only homogeneous chains, $s_k=s$, $\bar s_k=\bar s$, for all $k$.

\subsection{$L$ operators and monodromy matrices}
$L-$operators play a fundamental role in the theory of  integrable systems. In  the case of spin magnets they are defined as
follows
\begin{equation}\label{Lax}
L(u)=u+i\left(\begin{array}{cc}
S_0 & S_-\\
S_+& -S_0
\end{array}\right)\,, \qquad \bar L(\bar u)=\bar u+i\left(\begin{array}{cc}
\bar S_0 & \bar S_-\\
\bar S_+& -\bar S_0
\end{array}\right)\,.
\end{equation}
Here $u,\bar u$ are two complex numbers (spectral parameters). Note that  $L(u)\,\big(\bar L(\bar u)\big)$
acts on a tensor product of
$L_2(\mathbb{C})$ and a two dimensional complex vector space (auxiliary space),~$ \mathbb{V}_0\equiv \mathbb{C}^2$.
The operators $L(u)$ and $L'(v)$ acting on $L_2(\mathbb{C})\otimes \mathbb{V}_0$ and $L_2(\mathbb{C})\otimes
\mathbb{V}_{0'}$, respectively, satisfy the  fundamental commutation relation (FCR)
\begin{eqnarray}\label{FCR}
\mathcal{R}_{00'}(u-v) L(u) L'(v)&=& L'(v) L(u)\mathcal{R}_{00'}(u-v)\,,
\nonumber\\
\mathcal{R}_{00'}(\bar u-\bar v) \bar L(\bar u) \bar L'(\bar v)&=& \bar L'(\bar v) \bar L(\bar u)\mathcal{R}_{00'}(\bar u-\bar v)\,,
\end{eqnarray}
The operator $\mathcal{R}_{00'}(u)$  ($\mathcal{R}-$matrix) acts on the tensor product of two auxiliary spaces,
 $\mathbb{V}_0\otimes\mathbb{V}_{0'}=\mathbb{C}^2\otimes \mathbb{C}^2$,
and has the form $\mathcal{R}_{00'}(u)=u+i P_{00'}$ where $P_{00'}$ is the permutation operator on
$\mathbb{V}_0\otimes\mathbb{V}_{0'}$.
The monodromy matrix is defined  as a product of $L$ operators acting on the same auxiliary but different quantum
spaces
\begin{equation}\label{monodromy}
T(u)=L_1(u)L_2(u)\ldots L_N(u)\,, \qquad
 \bar T(\bar u)=\bar L_1(\bar u)\bar L_2(\bar u)\ldots \bar L_N(\bar u)\,.
\end{equation}
The $L-$operator with  subscript $k$ acts nontrivially  on the $k-$th space in the ternsor product~(\ref{HN}).
 The monodromy matrix $T_N(u)$  ( $\bar T_N(\bar u)$ ) is  a  two by two
matrix in  the auxiliary space with  entries that are  operators on  the quantum space $\mathbb{H}_N$
\begin{equation}
T(u)
=\left(\begin{array}{cc}
A_N(u)& B_N(u)\\
C_N(u)& D_N(u)
\end{array}\right)\,,
\qquad
\bar T(\bar u)
 =\left(\begin{array}{cc}
\bar A_N(\bar u)& \bar B_N(\bar u)\\
\bar C_N(\bar u)& \bar D_N(\bar u)
\end{array}\right)\,.
\end{equation}
Monodromy matrices satisfy the same commutation relation as $L-$operators, Eq.~(\ref{FCR})
\begin{eqnarray}\label{FCRT}
\mathcal{R}_{00'}(u-v) T_N(u) T'_N(v) & = &T'_N(v) T_N(u)\mathcal{R}_{00'}(u-v)\,,
\nonumber\\
\mathcal{R}_{00'}(\bar u-\bar v) \bar T_N(\bar u) \bar T'_N(\bar v) & = &\bar T'_N(\bar v) \bar T_N(\bar u)
\mathcal{R}_{00'}(\bar u-\bar v)\,.
\end{eqnarray}

These equations result in certain algebraic relations for the entries of the monodromy matrices.
In particular, they imply that  all operators commute with themselves for different values of the spectral parameter
\begin{eqnarray}\label{AA}
[A_N(u), A_N(v)]=0, \qquad [B_N(u), B_N(v)]=0, \nonumber\\ {}
 [C_N(u), C_N(v)]=0,\qquad [D_N(u), D_N(v)]=0
\end{eqnarray}
and similar for all others. By  construction the operators $A_N(u), D_N(u)$ are polynomials of degree $N$ in $u$,
while the operators $B_N(u), C_N(u)$ are  polynomials of a degree $N-1$,
\begin{eqnarray}\label{ABCDexp}
 A_N(u)=u^N+iu^{N-1}S_0+\sum_{k=2}^{N} u^{N-k} a_k\,, \qquad  B_N(u)=iS_-u^{N-1}+\sum_{k=2}^{N} u^{N-k} b_k\,,
\nonumber\\
 D_N(u)=u^N-iu^{N-1}S_0+\sum_{k=2}^{N} u^{N-k} d_k\,, \qquad  C_N(u)=iS_+u^{N-1}+\sum_{k=2}^{N} u^{N-k} c_k\,,
\end{eqnarray}
where $S_\alpha=\sum_{k=1}^N S_\alpha^{(k)}$ are the operators of  total spin. {
The construction for anti-holomorphic
sector is essentially the same and we will omit the corresponding similar expressions as a rule.}
It follows from (\ref{AA}),~(\ref{ABCDexp}) that $[S_0,a_k]=[a_i, a_k]=0$ for all $i,k$ and similar for $b_k,c_k,d_k$
operators. Taking into account that $A_N(u)^\dagger=\bar A_N(u^*)$ one concludes that the operators
\begin{equation}
i(S_0+\bar S_0), \qquad S_0-\bar S_0, \qquad  a^+_k=\frac12(a_k+\bar a_k), \qquad a^-_k=\frac{i}2(a_k-\bar a_k)\,,
\end{equation}
form a set of commuting self-adjoint  operators,
\begin{equation}
\mathcal{A}_N=\Big\{i(S_0+\bar S_0),\, S_0-\bar S_0,\, a^+_k,\,  a^-_k,\, \quad k=2,\ldots
N\Big\}
\end{equation}
and hence can be diagonalized simultaneously. We want to stress here that self-adjointness which does not play any
essential role in an analysis of finite-dimensional models  is very important in the case under
consideration~\footnote{Indeed, one can consider rotated monodromy matrices
$T'_N(u)=U T_N(u)U^{-1}, \bar T'_N(\bar u)=U \bar T_N(\bar u)U^{-1}$, where $U$ is a certain two by two matrix. The new entries
$A'_N(u), B'_N(u),\ldots$ obey all the same recurrence relations and form commutative families of operators.
However they are not self-adjoint and cannot be diagonalized.}.

The operators
$A_N(u), B_N(u),\ldots$ are differential operators of $N-$th order in the variables $z_1,\ldots,z_N$.
{Let $\Psi_A(\boldsymbol{z})=\Psi_A(z_1,\bar z_1\ldots,z_N,\bar z_N)$
be an eigenfunction of the operators $A_N(u),\bar A_N(\bar u)$.}
By virtue of Eq.~(\ref{ABCDexp})
 the corresponding eigenvalues are polynomials of degree $N$ in $u$, $\bar u$, respectively.
The eigenfunctions can be labelled by  zeroes of these polynomials,
i.e.
\begin{eqnarray}\label{AbAN}
A_N(u)\,\Psi_A ({\boldsymbol{x}}|\boldsymbol{z})&=&(u-x_1)\ldots(u-x_N)\,\Psi_A ({\boldsymbol{x}}|\boldsymbol{z})\,,
\nonumber\\
\bar A_N(\bar u)\,\Psi_A({\boldsymbol{x}}|\boldsymbol{z})&=&(\bar u-\bar x_1)\ldots(\bar u-\bar x_N)
\,\Psi_A ({\boldsymbol{x}}|\boldsymbol{z})\,,
\end{eqnarray}
where
\begin{eqnarray}\label{XZ}
{\boldsymbol{x}}=\{\boldsymbol{x}_1,
\ldots,\boldsymbol{x}_N
\}
, &\hskip 5mm &\boldsymbol{x}_k=(x_k,\bar x_k)
\nonumber\\
\boldsymbol{z}=\{\boldsymbol{z}_1,
\ldots,\boldsymbol{z}_N
\}, &\hskip 5mm& \boldsymbol{z}_k=(z_k,\bar z_k).
\end{eqnarray}
Note that a behaviour of the eigenfunction under the scale transformations, $z\to\lambda z$, is
controlled by the sum $i\sum_k x_k$ (which is the eigenvalue of the operator $S_0$)
\begin{equation}
\Psi_A ({\boldsymbol{x}}|\lambda\boldsymbol{z})=
\lambda^{-Ns+i\sum_k x_k}\bar\lambda^{-N\bar s+i\sum_k \bar x_k}\Psi_A ({\boldsymbol{x}}|\boldsymbol{z})\,.
\end{equation}

In full analogy with the previous case the operators $B_N, \bar B_N$ give rise to another set of the commuting operators,
\begin{equation}
\mathcal{B}_N=\Big\{i(S_-+\bar S_-),\ S_--\bar S_-,\ b^+_k=\frac12(b_k+\bar b_k),\
 b^-_k=\frac{i}2(b_k-\bar b_k), \  k=2,\ldots
N-1\Big\}.
\end{equation}
The   eigenfunctions  can be  parameterized by the momenta $p,\bar p$, which are the eigenvalues of the $S_-,\bar
S_-$ operators and  the roots of $x_k,\bar x_k$, $k=1,\ldots, N-1$ of the corresponding eigenvalues
\begin{eqnarray}
B_N(u)\,\Psi_B ({\boldsymbol{x}}|\boldsymbol{z})&=&p(u-x_1)\ldots(u-x_{N-1})\,\Psi_B ({\boldsymbol{x}}|\boldsymbol{z})\,,
\nonumber\\
\bar B_N(\bar u)\,\Psi_B ({\boldsymbol{x}}|\boldsymbol{z})&=&\bar p(\bar u-\bar x_1)\ldots(\bar u-\bar x_{N-1})\,
\Psi_B ({\boldsymbol{x}}|\boldsymbol{z})\,.
\end{eqnarray}
In order to keep  the same notations for the  $A$ and $B$ cases,
we  have put  $x_N=p, \bar x_N=\bar p$, i.e.
\begin{equation}
{\boldsymbol{x}}=\{\boldsymbol{x}_1,
\ldots,\boldsymbol{x}_{N-1},
\boldsymbol{x}_{N}=(p,\bar p)\}\,.
\end{equation}
It will be shown below that eigenfunctions of the operators $D_N$ and $C_N$  are related to those of $A_N$ and $B_N$ by an  inversion
transformation.
In  sect.~\ref{sect:iterative} we present an iterative procedure for  constructing the eigenfunctions.
It relies on the properties of  operators that factorize the general $\mathcal{R}-$matrix, which are discussed in the next
section.

\subsection{$\mathcal{R}$-matrix and factorizing operators}
General $\mathcal{R}-$matrix is defined as a solution of the $RLL-$relation~\cite{KRS}
\begin{eqnarray}\label{RLL}
\mathcal{R}_{\boldsymbol{s}_1\boldsymbol{s}_2}(u-v,\bar u-\bar v) L_{s_1}(u) L_{s_2}(v)&=&
L_{s_2}(v)L_{s_1}(u)\mathcal{R}_{\boldsymbol{s}_1\boldsymbol{s}_2}(u-v,\bar u-\bar v)\,,
\nonumber\\
\mathcal{R}_{\boldsymbol{s}_1\boldsymbol{s}_2}(u-v,\bar u-\bar v) \bar L_{\bar s_1}(\bar u) \bar L_{\bar s_2}(\bar v)&
=&
\bar L_{\bar s_2}(\bar v)\bar L_{\bar s_1}(u)\mathcal{R}_{\boldsymbol{s}_1\boldsymbol{s}_2}(u-v,\bar u-\bar v)\,.
\end{eqnarray}
Here $L-$operators act in the same auxiliary space but in different quantum spaces and  the operator
$\mathcal{R}_{\boldsymbol{s}_1\boldsymbol{s}_2}$
maps
$L_2(\mathbb{C})\otimes L_2(\mathbb{C})\mapsto L_2(\mathbb{C})\otimes L_2(\mathbb{C})$. The labels
$\boldsymbol{s}_k=(s_k, \bar s_k)$
indicate the representation  of the $SL(2,\mathbb{C})$ group in the first and  second quantum spaces. The
operator
$\mathcal{R}_{\boldsymbol{s}_1\boldsymbol{s}_2}$
satisfying  Eqs.~(\ref{RLL}) was  constructed  as an integral operator in~Ref.~\cite{Derkachov:2001yn}.
Later it has been suggested  to look for  the solutions of Eq.~(\ref{RLL}) in a factorized
form~\cite{Derkachov:2005hw}. Below we briefly describe the corresponding construction. First we note that
the
$L-$operator depends on two parameters: the spectral parameter~$\boldsymbol{u}$ and the spin $\boldsymbol{s}$.
It is convenient to
define two linear combinations~\footnote{ We will not display formulae for the anti-holomorphic sector
since they are identical to the ones  in holomorphic sector.}
\begin{equation}\label{u1u2}
u_1=u-i(1-s)\,, \qquad u_2=u-is.
\end{equation}
Thus $L_{s_1}(u)=L(u_1,u_2)$ and $ L_{s_2}(v)=L(v_1,v_2)$. Factoring out the permutation operator from
$\mathcal{R}-$matrix,\
$\mathcal{R}_{\boldsymbol{s}_1\boldsymbol{s}_2}=P_{12} \widehat {\mathcal{R}}_{12}$,\
one gets the following equation on $\widehat {\mathcal{R}}_{12}$
\begin{equation}\label{PRLL}
\widehat {\mathcal{R}}_{12}\, L_1(u_1,u_2) L_2(v_1,v_2)=L_1(v_1,v_2) L_2(u_1,u_2)\,\widehat {\mathcal{R}}_{12}\,.
\end{equation}
The operator $L_1 (L_2)$ acts on the first (second) space in the tensor product, $L_2(\mathbb{C})\otimes
L_2(\mathbb{C})$ (i.e. $L_1$ and $L_2$ are the differential operators in $z_1$ and $z_2$, respectively.)
Thus the operator $\widehat {\mathcal{R}}_{12}$ interchanges the parameters
$(u_1,u_2)\leftrightarrow(v_1,v_2)$ in the product of two $L$ operators. It is natural to break this
permutation of the parameters into two operations and construct the operators
 which
interchange the parameters $u_1\leftrightarrow v_1$ and $u_2\leftrightarrow v_2$
in the product of $L-$operators separately
\begin{eqnarray}\label{R12}
{\mathcal{R}^{(1)}_{12}}\, L_1(u_1,u_2) L_2(v_1,v_2)&=&L_1(v_1,u_2) L_2(u_1,v_2)\,{\mathcal{R}^{(1)}_{12}}\,,
\nonumber\\
{\mathcal{R}^{(2)}_{12}}\, L_1(u_1,u_2) L_2(v_1,v_2)&=&L_1(u_1,v_2) L_2(v_1,u_2)\,{\mathcal{R}^{(2)}_{12}}\,.
\end{eqnarray}
It turns out that the operators $\mathcal{R}^{(a)}_{12}$ depend only on the specific combinations
of  the spectral parameters~\footnote{
In order to avoid misunderstanding we stress that the factorizing operators $\mathcal{R}_{12}^{(a)}$ depend also on the
anti-holomorphic spectral parameters, i.e.
$
\mathcal{R}^{(1)}_{12}=\mathcal{R}^{(1)}_{12}(u_1-v_1,\bar u_1-\bar v_1;  u_1-v_2, \bar u_1-\bar v_2),
$
and satisfy the exchange relations~(\ref{R12}) with anti-holomorphic $L$ operators,
($L_1(u_1,u_2)\to \bar L_1(\bar u_1,\bar u_2)$, etc.).
}
\begin{equation}
\mathcal{R}^{(1)}_{12}= \mathcal{R}^{(1)}_{12}(u_1-v_1, u_1-v_2), \qquad
\mathcal{R}^{(2)}_{12} = \mathcal{R}^{(2)}_{12}(u_1-v_2, u_2-v_2)
\end{equation}
and  have a remarkably simple form~\cite{Derkachov:2005hw,Derkachov:2010zz}
\begin{eqnarray}\fl
[\mathcal{R}^{(1)}_{12}(u_1-v_1, u_1-v_2)\Phi](z_1,z_2)&=&\int d^2 w_2 \,
 \frac{[z_2-z_1]^{i(v_1-v_2)}}{[z_2-w_2]^{1-i(u_1-v_1)}[z_1-w_2]^{i(u_1-v_2)}}
\,\Phi(z_1,w_2)\,,
\nonumber\\[2mm]
\fl
[\mathcal{R}^{(2)}_{12}(u_1-v_2, u_2-v_2)\Phi](z_1,z_2)&=&
\int d^2 w_1 \, \frac{[z_1-z_2]^{i(u_1-u_2)}}{[w_1-z_1]^{1-i(u_2-v_2)}[w_1-z_2]^{i(u_1-v_2)}}
\,\Phi(w_1,z_2)\,, \label{Rint}
\end{eqnarray}
where $[a]^{\alpha}\equiv a^{\alpha}\, \bar a^{\bar\alpha}$ which is a single valued function in the complex plane provided that
$\alpha-\bar\alpha\in \mathbb{Z}$. The requirement of single-valuedness of the kernels results in quantization of the
spectral parameters, $u,\bar u$, $u-\bar u\in \mathbb{Z}$~\cite{Derkachov:2001yn}, which were so far considered as independent
variables.

Finally, the $\mathcal{R}-$matrix satisfying $RLL-$relation~(\ref{RLL}) is constructed as follows
\begin{equation}\label{Rmatrix}
\mathcal{R}_{\boldsymbol{s}_1\boldsymbol{s}_2} (u-v,\bar u-\bar v)=
P_{12}\mathcal{R}^{(1)}_{12}(u_1-v_1, u_1-u_2)\mathcal{R}^{(2)}_{12}(u_1-v_2, u_2-v_2)\,.
\end{equation}
For a  more detailed  discussion of properties of factorizing operators see Ref.~\cite{Derkachov:2006fw,Derkachov:2010zz}.

\section{Iterative construction of eigenfunctions}\label{sect:iterative}

We  present in this section a recurrence procedure of construction the eigenfunctions of the operators $A_N(u)$, $B_N(u)$.
(For simplicity we consider the homogeneous spin chain $s_k=s, \bar s_k=\bar s$ though the construction are easily
generalized  for general case.)

Let us consider a modified monodromy matrix
\begin{equation}
{T}_N(u,v)=L_1(u_1,v)L_2(u_1,u_2)\ldots L_N(u_1,u_2)\,.
\end{equation}
Here all $L-$operators except the first one has a standard form ($L_k(u)=L_k(u_1,u_2)$) while in the  first one
we replace the parameter $u_2=u-is\to v$. Taking into account that the first row of the $L-$operators does not change
under this substitution (see Eqs.~(\ref{Lax}),(\ref{u1u2}))
\begin{equation}
L_1(u_1,v_1)=\left(\begin{array}{cc}
u+iS^{(1)}_0 & S_-^{(1)}\\
\star &\star
\end{array}\right)
\end{equation}
one immediately gets  that such a modification leaves the elements in the  first row of the monodromy matrix intact,
\begin{equation}
{T}_N(u,v)=\left(\begin{array}{cc}
A_N(u) & B_N(u)\\
\star &\star
\end{array}\right)\,.
\end{equation}
Let us consider the commutation relation of the monodromy matrix ${T}_N(u,v)$ with an operator $\boldsymbol{\Lambda}_N(u,v)$
defined by~\footnote{Let us repeat here that we do not display  explicitly  the dependence on anti-holomorphic parameters,
that is
$\boldsymbol{\Lambda}_N(u,v)\equiv \boldsymbol{\Lambda}_N(\{u,\bar u\},\{v,\bar v\})$.}
\begin{equation}
\boldsymbol{\Lambda}_N(u,v)=\mathcal{R}^{(2)}_{12}(u_1-v, u_2-v)\mathcal{R}^{(2)}_{23}(u_1-v, u_2-v)\ldots\mathcal{R}^{(2)}_{N-1,N}(u_1-v, u_2-v)\,.
\end{equation}
Taking into account the relations~(\ref{R12}) one  obtains
\begin{equation}\label{TTp}
{T}_N(u,v)\,\boldsymbol{\Lambda}_N(u,v)=\boldsymbol{\Lambda}_N(u,v)\,{T}_{N-1}(u)\,L_N(u_1,v)\,,
\end{equation}
where ${T}_{N-1}(u)=L_1(u)\ldots L_{N-1}(u)$. Comparing the matrix elements  in the first row of the l.h.s and r.h.s
of Eq.~(\ref{TTp})
one gets
\begin{eqnarray}\label{ABN}\fl
A_N(u)\boldsymbol{\Lambda}_N(u,v)&=&\boldsymbol{\Lambda}_N(u,v) \Big(A_{N-1}(u)(u+is+iz_N\partial_{z_N})
+B_{N-1}(u)z_N(iz_N\partial_{z_N}+
v-u_1+i)\Big)\,,
\nonumber\\
\fl
B_N(u)\boldsymbol{\Lambda}_N(u,v)&=&\boldsymbol{\Lambda}_N(u,v) \Big(B_{N-1}(u)(v-iz_N\partial_{z_N})-A_{N-1}(u)\partial_{z_N}\Big)\,.
\end{eqnarray}
%
%

\subsection{$B-$system}
\begin{center}
\begin{figure}[t]
\centerline{\includegraphics[width=0.4\linewidth]{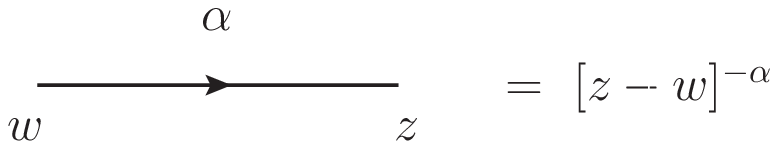}}
\caption{The diagrammatic  representation of the propagator.}
\label{Propagator}
\end{figure}
\end{center}
Let us apply the operators on both sides of the second of Eqs.~(\ref{ABN}) to the function
$\Psi(z_1,\ldots,z_{N-1})$ which does not depend on the variable $z_N$. In this case the second term on
the
r.h.s. ($A_{N-1}(u)\partial_{z_N}$) vanishes and the equation takes the form
\begin{equation}\label{BBL}
B_N(u)\,\boldsymbol{\Lambda}_N(u,v)\,
\Psi(z_1,\ldots,z_{N-1})=v\,\boldsymbol{\Lambda}_N(u,v)\, B_{N-1}(u)\,\Psi(z_1,\ldots,z_{N-1})\,,
\end{equation}
so that the operator $\boldsymbol{\Lambda}_N(u,v)$ intertwines the operators $B_{N-1}(u)$ and $B_{N}(u)$.
It is useful to rewrite Eq.~(\ref{BBL}) in an operator form
\begin{equation}\label{BL}
B_N(u)\,\Lambda_N{(x,\bar x)}=(u-x)\,\Lambda_N{(x,\bar x)}\,
 B_{N-1}(u)\,,
\end{equation}
where the operator  $\Lambda_N{(x,\bar x)}$ maps functions of $N-1$ variables $z_1,\ldots, z_{N-1}$ to the
functions of $N$ variables $z_1,\ldots, z_N$ and is defined as follows
\begin{equation}\label{LL1}
\Lambda_N{(x,\bar x)}\Psi(z_1,\ldots,z_{N-1}) =r_N(x,\bar x)\,\boldsymbol{\Lambda}_N(u,v)\Big|_{v=u-x,\bar v=\bar u -\bar x}\Psi(z_1,\ldots,z_{N-1})\,.
\end{equation}
It can be easily checked that for this choice of the parameters $v,\bar v$  the r.h.s. of Eq.~(\ref{LL1})
depends only on $ \boldsymbol{x}=(x,\bar x)$ and does not depend on the spectral parameters $u,\bar u$.

\begin{figure}[t]
\centerline{\includegraphics[width=0.8\linewidth]{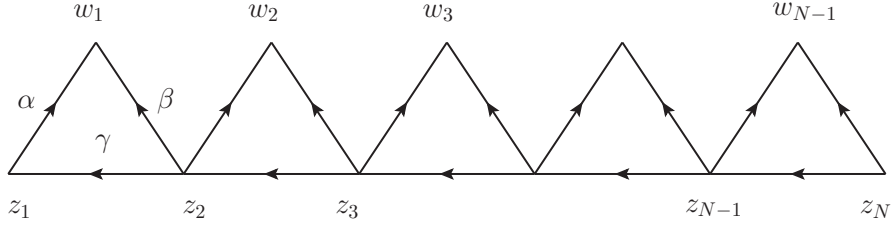}}
\caption{The diagrammatic representation for the kernel $\Lambda_N^{(x_1,\bar x_1)}(z_1,\ldots,z_N|w_1,\ldots,w_{N-1})$
(up to factor $r_N(x,\bar x)$).
The arrow with index $\alpha$ from $z$ to $w$ stands for $(w-z)^{-\alpha}(\bar w-\bar z)^{\bar \alpha}$.
The indices are given by the following expressions: $\alpha=1-s-ix$, $\beta=1-s+ix$, $\gamma=2s-1$.}
\label{LambdaB}
\end{figure}

Making use of Eq.~(\ref{Rint}) one can represent the operator $\Lambda_N{(\boldsymbol{x})}$ as an integral operator.
Its kernel in the diagrammatic form is shown
in  Fig.~\ref{LambdaB}.
It has the form of a  Feynman diagram, see Fig.~\ref{Propagator}, where an arrow from the point $w$ to $z$ and index $\alpha$ stands for the
``propagator'',
\begin{equation}
G_\alpha(z-w)=(z-w)^{-\alpha}(\bar z-\bar w)^{-\bar\alpha}\equiv[z-w]^{-\alpha}\,.
\end{equation}

Here we introduced a short-hand notation, $[z]^{\alpha}=z^{\alpha}\,\bar z^{\bar\alpha}$.
It is convenient to choose the normalization factor $r_N(x,\bar x)$  as follows
\begin{equation}
r_N(x,\bar x)=(a(s+ix) a(\bar s-i\bar x))^{N-1}\,, \qquad a(\alpha)=\frac{\Gamma(1-\bar\alpha)}{\Gamma(\alpha)}\,.
\end{equation}
For this choice of $r_N(x)$ the operators $\Lambda_N$ and $\Lambda_{N-1}$ satisfy the exchange relation
\begin{equation}\label{exchange-R}
\Lambda_N{(\boldsymbol{x}_1)}\,\Lambda_{N-1}{(\boldsymbol{x}_2)}=\Lambda_N{(\boldsymbol{x}_2)}\,\Lambda_{N-1}{(\boldsymbol{x}_1)}\,
\end{equation}
which can be proven with  the help of the diagram technique developed in Ref.~\cite{Derkachov:2001yn}.
%

Now it is easy to see that the eigenfunctions of the
operators $B_N(u)$ and $\bar B_N(\bar u)$ have the form
\begin{equation}\label{defPhi}
\Psi_B ({\boldsymbol{x}}|\boldsymbol{z})=|p|^{N-1}\,
\Lambda_N{(\boldsymbol{x}_1)}\,\Lambda_{N-1}{(\boldsymbol{x}_2)}
\ldots \Lambda_2{(\boldsymbol{x}_{N-1})}\, e^{ip  z_1+i\bar p \bar z_1}\,.
\end{equation}
Each operator $\Lambda_k{(\boldsymbol{x}_k)}$ maps a function of $k-1$ variables to a function of $k$ variables.
Thus the product of the operators in (\ref{defPhi}) maps the function of one variable, $e^{ip  z_1+i\bar p \bar z_1}$,
to a function of $N-$variables.
In order to obtain the  conventional normalization of the eigenfunctions  we included the prefactor
$|p|^{N-1}$ in the  definition~(\ref{defPhi}).
Taking into account Eq.~(\ref{BL})  and  using that $B_1(u)=S_1=-\partial_{z_1}$ one obtains
\begin{equation}\fl
B_N(u)\,\Psi_B ({\boldsymbol{x}}|\boldsymbol{z})=p\prod_{k=1}^{N-1} (u-x_k) \,
\Psi_B ({\boldsymbol{x}}|\boldsymbol{z})\,,
\qquad
\bar B_N(\bar u)\,\Psi_B ({\boldsymbol{x}}|\boldsymbol{z})=
\bar p\prod_{k=1}^{N-1} (\bar u-\bar x_k) \,\Psi_B ({\boldsymbol{x}}|\boldsymbol{z})\,.
\end{equation}
Note also that due to the exchange relation~(\ref{exchange-R}) the eigenfunctions
are
symmetric functions of  the parameters $(\boldsymbol{x_1},\ldots\boldsymbol{x}_{N-1})$.

Let us  figure out which are the possible values  of the  variables $(\boldsymbol{x_1},\ldots\boldsymbol{x}_{N-1})$..
By  construction  the  variables  $(\{x_1,\bar x_1\},\ldots\{x_{N-1},\bar
x_{N-1}\})$ satisfy the restriction
\begin{equation}\label{xbarxk}
(s+ix_k)-(\bar s-i\bar x_k)=i(x_k-\bar x_k)+n_s\in \mathbb{Z}
\end{equation}
for all $k$.
Further, the operator  $B_N(u)$ is a hermitian adjoint of $\bar B_N(\bar u)$,
$\bar B_{N}(\bar u)=(B_N(u))^\dagger$, provided that $u^*=\bar u$. It  results in  the following relation for the eigenvalues
\begin{equation*}
\left(\prod_{k=1}^{N-1} (u-x_k)\right)^*=\prod_{k=1}^{N-1} (u^*-\bar x_k)
\end{equation*}
that, in its turn,  implies that $x_k^*=\bar x_k $. Together with  the condition~(\ref{xbarxk}) it results in the following
parametrization~\cite{Derkachov:2001yn}
\begin{equation}\label{x-n-nu}
x_k=-\frac{in_k}{2}+\nu_k\,, \qquad \bar x_k=\frac{in_k}{2}+\nu_k\,,
\end{equation}
where $\nu_k$ is real and  $n_k$ is integer (if $n_s$ is integer) or half-integer (if $n_s$ is half-integer).

\subsection{$A-$system}
\begin{figure}[t]
\centerline{\includegraphics[width=0.85\linewidth]{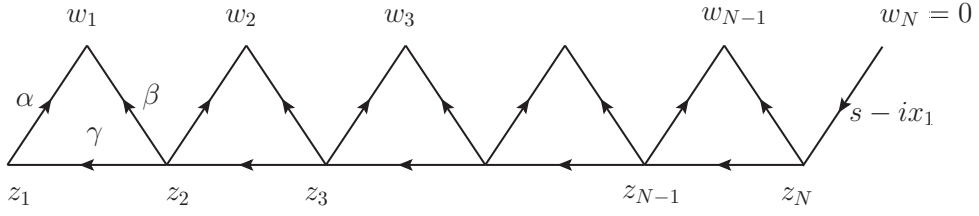}}
\caption{The diagrammatic representation for the kernel $\tilde \Lambda_N^{(x_1,\bar x_1)}(z_1,\ldots,z_N|w_1,\ldots,w_{N-1})$
(up to factor $r_N(x,\bar x)$).
The indices are given by the following expressions: $\alpha=1-s-ix$, $\beta=1-s+ix$, $\gamma=2s-1$.}
\label{LambdaA}
\end{figure}

The construction of the eigenfunctions of the operator $A_N(u)$ goes along the same lines.
Let us apply both sides of the first of  Eqs.~(\ref{ABN})
to a function $\Psi$ which depends on $z_N$ and $\bar z_N$ in the specific way, $\Psi=[z_N]^{i(u_1-v)-1}
\Psi_{N-1}(z_1,\ldots, z_{N-1})$.
The second term ($\sim B_{N-1}(u)$) on the r.h.s. of this equation
vanishes so that one obtains
\begin{equation}\label{AF}\fl
A_N(u)\boldsymbol{\Lambda}_N(u,v)[z_N]^{i(u_1-v)-1}\Psi_{N-1}
=(u+is-u_1+v-i)\boldsymbol{\Lambda}_N(u,v) A_{N-1}(u)[z_N]^{i(u_1-v)-1}\Psi_{N-1}\,.
\end{equation}
Taking into account explicit expression for the operator $\mathcal{R}_{N-1,N}^{(2)}$, Eq.~(\ref{R12}), it is easy to
verify that $\boldsymbol{\Lambda}_N(u,v) z_N= z_N\boldsymbol{\Lambda}_N(u,v)$. Finally, substituting
$v=u-x$ and multiplying both sides of (\ref{AF}) by the normalization factor $r_N(x,\bar x)$ one obtains
\begin{equation}
A_N(u)\, \tilde \Lambda_N{(x,\bar x)}
=(u-x)\,\tilde \Lambda_N{(x,\bar x)}\, A_{N-1}(u)\,.
\end{equation}
The operator
\begin{equation}
\tilde \Lambda_N{(\boldsymbol{x})}=[z_N^{ix-s}]\,\Lambda_N{(\boldsymbol{x})}\equiv z_N^{ix-s}\,\bar z_N^{i\bar x-\bar s}\,\Lambda_N{(\boldsymbol{x})}
\end{equation}
 maps a function of  $N-1$ variables
to a function
of $N-$variables.
The diagrammatic representation for the kernel of the operator $\tilde \Lambda_N{(\boldsymbol{x}_1)}$ is shown in
Fig.~\ref{LambdaA}.

The eigenfunctions of the operators $A_N(u)$, $\bar A_N(\bar u)$ are constructed using  the same scheme as the
eigenfunctions
of $B_N$ operators. Namely,
\begin{equation}\label{PsiA}
\Psi_A ({\boldsymbol{x}}|\boldsymbol{z})=
\tilde \Lambda_N{(\boldsymbol{x}_1)}\
\ldots \tilde \Lambda_{2}{(\boldsymbol{x}_{N-1})} \tilde \Lambda_{1}{(\boldsymbol{x}_N)}=
\tilde \Lambda_N{(\boldsymbol{x}_1)}\,
\ldots \tilde \Lambda_{2}{(\boldsymbol{x}_{N-1})} [z_1]^{ix_1-s}\,.
\end{equation}
The diagrammatic representation for $\Psi_A ({\boldsymbol{x}}|\boldsymbol{z})$ is shown in Fig.~\ref{LambdaA}.
Evidently this function satisfies Eqs.~(\ref{AbAN}). The eigenfunction~(\ref{PsiA}) is symmetric
under permutations of  variables,
$\boldsymbol{x}_k\leftrightarrow \boldsymbol{x}_j$. This property follows from the
exchange relation
\begin{equation}\label{exchange-A}
\tilde \Lambda_N{(\boldsymbol{x}_1)}\,\tilde \Lambda_{N-1}{(\boldsymbol{x}_2)}=\tilde \Lambda_N{(\boldsymbol{x}_2)}\,
\tilde \Lambda_{N-1}{(\boldsymbol{x}_1)}\,
\end{equation}
which can  be proven using the same diagrammatic technique.

\subsection{$C$ and $D$ systems}
The eigenfunctions of the operators $D_N$ and $C_N$  are related to those of $A_N$ and $B_N$ by an  inversion
transformation. The inversion operator $J$,
\begin{equation}
[{J} \varphi](z_1,\ldots,z_N)=\psi(z_1,\ldots,z_N)=
\prod_{k=1}^{N}z_k^{-2s}\bar z_k^{-2\bar s}\varphi\left(\frac1{z_1},\ldots,\frac1{z_N}\right)\,,
\end{equation}
generates the following transformation of  the $sl(2)$ algebra
\begin{equation}
{J} \,S^{(k)}_\pm\, {J}=S^{(k)}_\mp\,, \quad {J}\, S^{(k)}_0\,{J}=-S^{(k)}_0\,, \quad
{J} \,\bar S^{(k)}_\pm\, {J}=\bar S^{(k)}_\mp\,, \quad {J}\, \bar S^{(k)}_0\,{J}=-\bar S^{(k)}_0\,.
\end{equation}
The $L-$operators (and hence monodromy matrices) transform  under the inversion as follows
%
%
%
\begin{eqnarray}
{J}\, L_k(u)\, {J}=\sigma_1 \,L_k(u)\,\sigma_1\,, &\phantom{bbbb}&
{J}\, \bar L_k(\bar u)\, {J}=\sigma_1 \,\bar L_k(\bar u)\,\sigma_1\,,
\nonumber\\
\label{TI}
{J}\, T_N(u)\, {J}=\sigma_1 \,T_N(u)\,\sigma_1\,, &\phantom{bbbb}&
{J}\, \bar T_N(\bar u)\, {J}=\sigma_1 \,\bar T_N(\bar u)\,\sigma_1\,,
\end{eqnarray}
%
where $\sigma_1$ is the Pauli matrix. From Eqs.~(\ref{TI}) one  immediately derives
\begin{eqnarray}
{J}\, A_N(u)=D_N(u)\,{J}\,, &\phantom{bbbb}& {J}\, \bar A_N(\bar u)=\bar D_N(\bar u)\,{J}\,,
\nonumber\\
{J}\, B_N(u)=C_N(u)\,{J}\,, &\phantom{bbbb}& {J}\, \bar B_N(\bar u)=\bar C_N(\bar u)\,{J}\,.
\end{eqnarray}
Thus the eigenfunctions of the $D_N(u),\bar D_N(\bar u)$ ($C_N(u),\bar C_N(\bar u)$)
commutative family  are related to those of $A_N(u),\bar A_N(\bar u)$ ($B_N(u),\bar B_N(\bar u)$)
by inversion. Namely, for the $D-$system one obtains
\begin{equation}\label{PsiD}\fl
\Psi_D ({\boldsymbol{x}}|\boldsymbol{z})= {J}\,\Psi_A({\boldsymbol{x}}|\boldsymbol{z})=
\widehat \Lambda_N{(x_1,\bar x_1)}\,\widehat \Lambda_{N-1}{(x_2,\bar x_2)}
\ldots \widehat \Lambda_{2}{(x_{N-1},\bar x_{N-1})}\,\widehat \Lambda_{1}{(x_N,\bar x_N)}\,,
\end{equation}
where $\widehat \Lambda_k{(x,\bar x)}=z_1^{-ix-s}\,\bar z_1^{-ix-s}\,\Lambda_k{(x,\bar x)}$.
In turn, for the $C-$system one gets
\begin{eqnarray}\fl
\Psi_C({\boldsymbol{x}}|\boldsymbol{z})&=& {J}\,\Psi_B ({\boldsymbol{x}}|\boldsymbol{z})
\nonumber\\
\fl
&=&|p|^{N-1}\,
\bar \Lambda_N{(x_1,\bar x_1)}\,\bar \Lambda_{N-1}{(x_2,\bar x_2)}
\ldots \bar \Lambda_{2}{(x_{N-1},\bar x_{N-1})} z_1^{-2s} \bar z_1^{-2\bar s} e^{ip/z_1+i\bar p/\bar z_1}\,,
\end{eqnarray}
with $\bar \Lambda_k{(x,\bar x)}=z_1^{-ix-s}\,\bar z_1^{-i\bar x-\bar s}\,z_k^{ix-s}\,\bar z_k^{i\bar x-\bar s}\,
\Lambda_k{(x,\bar x)}$.

\begin{figure}[t]
\centerline{\includegraphics[width=0.7\linewidth]{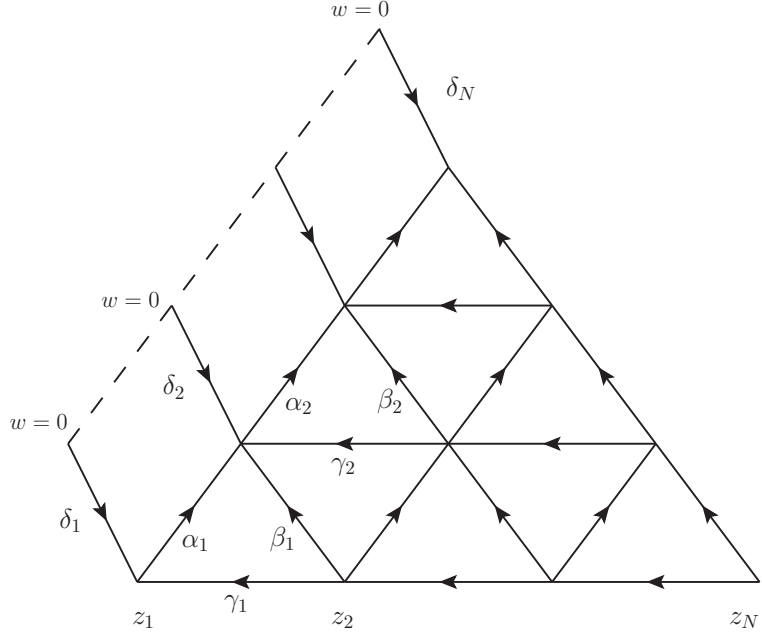}}
\caption{The diagrammatic representation of the eigenfunction $\Psi_D(\boldsymbol{x}|\boldsymbol{z})$.
Here $\alpha_k=1-s-ix_k$, $\beta_k=1-s+ix_k$, $\gamma_k=2s-1$ and $\delta_k=s+ix_k$. The dashed line stands for the point  $w=0$.
}
\label{fig:PsiD}
\end{figure}

\section{Scalar products and Sklyanin's measure}\label{sklyanin-measure}

The functions $\Psi_{S}({\boldsymbol{x}}|\boldsymbol{z})$, $S=A,B,C,D$,
being eigenfunctions of the self-adjoint operators form a complete orthonormal basis in the Hilbert space $\mathbb{H}_N$.
Arbitrary function $\Phi\in \mathbb{H}_N$ can be expanded in this basis as follows
\begin{equation}
\Phi(z)=\int \mathcal{D}_S\boldsymbol{x}\, \boldsymbol{ \mu}_{S}(\boldsymbol{x})\,
C_S(\boldsymbol{x})\,\Psi_{S} ({\boldsymbol{x}}|\boldsymbol{z})\,.
\end{equation}
The symbol $\mathcal{D}_S\boldsymbol{x}$ stands for
\begin{equation}
\mathcal{D}_{A(D)}\boldsymbol{x}=\prod_{k=1}^N \left(\sum_{n_k=-\infty}^{\infty}\int_{-\infty}^{\infty} d\nu_k\right)\,, \quad
\mathcal{D}_{B(C)}\boldsymbol{x}=d^2p\,\prod_{k=1}^{N-1} \left(\sum_{n_k=-\infty}^{\infty}\int_{-\infty}^{\infty} d\nu_k\right)\,.
\end{equation}
Depending on the value of the spin in the quantum space, $n_s=s-\bar s$,
the sum over $n_k$ goes over all integers (integer $n_s$)  or half-integers (half-integer $n_s$).  The weight function
$\boldsymbol{ \mu}_S(\boldsymbol{x})$ is the so-called
Sklyanin's measure and the function $C_S(\boldsymbol{x})$ is given by the scalar product
\begin{equation}
C_S(\boldsymbol{x})=\vev{\Psi_{S} ({\boldsymbol{x}}|\boldsymbol{z})|\Phi(\boldsymbol{z})}\,.
\end{equation}
Sklyanin's measure $\boldsymbol{\mu}_S(\boldsymbol{x})$ is related to the scalar product of the eigenfunctions
\begin{equation}\label{SCS}
\vev{\Psi_{S} ({\boldsymbol{x}'}|\boldsymbol{z})|\Psi_{S} ({\boldsymbol{x}}|\boldsymbol{z})}=
 \boldsymbol{ \mu}^{-1}_{S}(\boldsymbol{x})\,\delta_S(\boldsymbol{x}-\boldsymbol{x}')\,.
\end{equation}
Here the delta function $\delta_S(\boldsymbol{x}-\boldsymbol{x}')$ is defined as follows:
\begin{itemize}
\item
For $S=A,D$
\begin{equation}
\delta_S(\boldsymbol{x}-\boldsymbol{x}')=\frac1{N!}
\sum_{S_N}\delta(\boldsymbol{x}_1-\boldsymbol{x}'_{k_1})\ldots
\delta(\boldsymbol{x}_N-\boldsymbol{x}'_{k_N})\,.
\end{equation}

\item For $S=B, C$
\begin{equation}
\delta_S(\boldsymbol{x}-\boldsymbol{x}')=\frac1{(N-1)!}\delta^2(\vec{p}-\vec{p}')
\sum_{S_{N-1}}\delta(\boldsymbol{x}_1-\boldsymbol{x}'_{k_1})\ldots
\delta(\boldsymbol{x}_{N-1}-\boldsymbol{x}'_{k_{N-1}})\,.
\end{equation}
\end{itemize}
In above expressions summation goes over all permutations of $N$ and $N-1$ elements, respectively and
\begin{equation}
\delta(\boldsymbol{x}_k-\boldsymbol{x}'_m)\equiv  \delta_{n_kn'_m} \delta(\nu_k-\nu'_m)
\end{equation}

The calculation of the scalar product~(\ref{SCS}) is based on the following exchange relations for $\Lambda$
operators
%
%
%
\numparts
\begin{eqnarray}\label{exchangeB}
\Lambda_k^\dagger(\boldsymbol{x}'_k)\,\Lambda_k(\boldsymbol{x}_k)& = &
\alpha(\boldsymbol{x}_k,\boldsymbol{x}'_k)\,
\Lambda_{k-1}(\boldsymbol{x}_k)\Lambda_{k-1}^\dagger(\boldsymbol{x}'_k)\,,
\\[2mm]
\label{exchangeA}
\tilde \Lambda_k^\dagger(\boldsymbol{x}'_k)\,\tilde \Lambda_k(\boldsymbol{x}_k)& = &
\alpha(\boldsymbol{x}_k,\boldsymbol{x}'_k)\,
\tilde \Lambda_{k-1}(\boldsymbol{x}_k)\tilde \Lambda_{k-1}^\dagger(\boldsymbol{x}'_k)\,,
\end{eqnarray}
\endnumparts
%
where it is assumed that $\boldsymbol{x}'_k\neq \boldsymbol{x}_k$ and
\begin{equation}\label{alphaxx}
\alpha(\boldsymbol{x}_k,\boldsymbol{x}'_k)=\frac{\pi^2}{(x_k-x'_k)(\bar x_k-\bar x'_k)}\,.
\end{equation}

\begin{figure}[t]
\centerline{\includegraphics[width=0.95\linewidth]{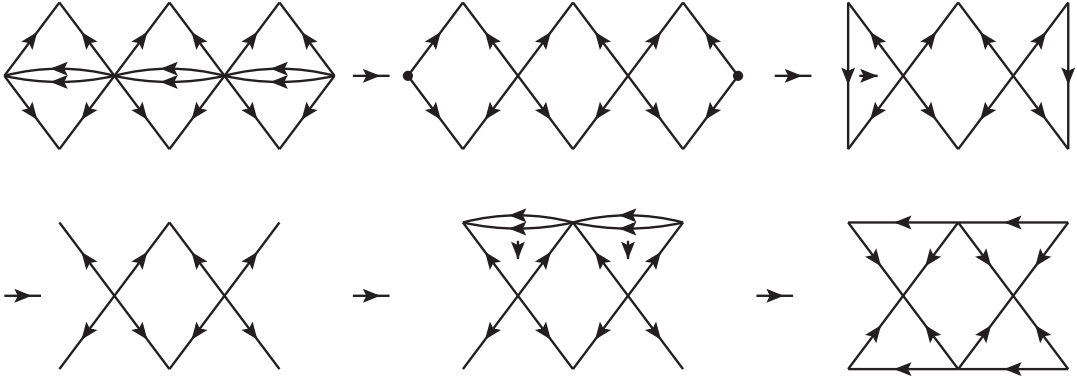}}
\caption{An illustration to the diagrammatic proof of the exchange relation~(\ref{exchangeB}).}
\label{ProofB}
\end{figure}

The   relations~(\ref{exchangeB}), (\ref{exchangeA}) can be proven diagrammatically.
Namely, one can show that the diagrams on the l.h.s and r.h.s. of Eq.~(\ref{exchangeA}),(\ref{exchangeB})  can be brought to  the same form
after a certain sequence  of transformations. The transformations relevant for Eq.~(\ref{exchangeB})
are  shown schematically in Fig.~\ref{ProofB}.
This technique and its application to the analysis of the $SL(2,\mathbb{C})$ spin chains was discussed at length
in Ref.~\cite{Derkachov:2001yn}. Therefore we will not go in much detail here and  only
comment briefly on the sequence of  transformations shown in Fig.~\ref{ProofB}.
 i) The right-most diagram in the first row is a diagrammatic representation for
the kernel $\Lambda_k^\dagger(\boldsymbol{x}'_k)\,\Lambda_k(\boldsymbol{x}_k)$ ($k=3$). The lines inside the rhombuses have
indices
$2s-1$ and $1-2s$ and therefore cancel (their product is equal to $1$). ii)~One integrates over the right-most and leftmost vertices
using the chain integration rule~(\ref{Chain}). iii)~The left-most vertical line is moved with the help of the cross
identity~(\ref{Cross}) to the right where it cancels with leftmost vertical line resulting in the first diagram in
the second line iv) One  inserts unity given by the product of two  lines with indices $(1-2s)$ (upper line) and
$(2s-1)$ (lower line) and moves lower lines down using cross identity~(\ref{Cross}). v) One flips arrows on the
lines (except the horizontal  ones). The last diagram in the second row coincides up to prefactor with a diagram for
the kernel of the product of the opartors
$\Lambda_{k-1}(\boldsymbol{x}_k)\Lambda_{k-1}^\dagger(\boldsymbol{x}'_k)$ ($k=3$). Collecting all factors which arise during
these transformations one arrives at Eq.~(\ref{exchangeB}).

The proof of the second relation, Eq.~(\ref{exchangeA}), goes along the same lines and we will not discuss~it.
\vskip 3mm

Let us come back to the  calculation of the scalar product~(\ref{SCS}).
Using the representations~(\ref{defPhi}) and (\ref{PsiA}) for the eigenfunctions
and making use of the
 exchange relations~(\ref{exchangeA}),~(\ref{exchangeB})
one can bring the scalar product~(\ref{SCS}) to the form
%
\numparts
\begin{eqnarray}\label{scA}\fl
\vev{\Psi_{A} ({\boldsymbol{x}'}|\boldsymbol{z})|\Psi_{A} ({\boldsymbol{x}}|\boldsymbol{z})}&=&
M_A(\boldsymbol{x},\boldsymbol{x}')\,\tilde \Lambda_1^\dagger(\boldsymbol{x}'_{N})\tilde \Lambda_1(\boldsymbol{x}_1)\,
\tilde \Lambda_1^\dagger(\boldsymbol{x}'_{N-1})\tilde \Lambda_1(\boldsymbol{x}_2)\ldots
\tilde \Lambda_1^\dagger(\boldsymbol{x}'_{1})\tilde \Lambda_1(\boldsymbol{x}_{N})\,,
\\[2mm]\fl
\vev{\Psi_{B} ({\boldsymbol{x}'}|\boldsymbol{z})|\Psi_{B} ({\boldsymbol{x}}|\boldsymbol{z})}&=&
M_B(\boldsymbol{x},\boldsymbol{x}')\,
\nonumber\\
\label{scB}
&\times&
\langle E_{p'}(z) \big|\Lambda_2^\dagger(\boldsymbol{x}'_{N-1}) \Lambda_2(\boldsymbol{x}_1)\,
\Lambda_2^\dagger(\boldsymbol{x}'_{N-2}) \Lambda_2(\boldsymbol{x}_2)\ldots
\Lambda_2^\dagger(\boldsymbol{x}'_{1}) \Lambda_2(\boldsymbol{x}_{N-1})\big| E_p(z)\rangle\,,
\end{eqnarray}
\endnumparts
%
where
$
E_p(z)=e^{ip  z+i\bar p \bar z}.
$
 In order to use the exchange relations
one  has to  assume that $\boldsymbol{x}'_{N-k}\neq \boldsymbol{x}_j$ for $j\neq k$ in the  product~(\ref{scA})
and $\boldsymbol{x}'_{N-1-k}\neq \boldsymbol{x}_j$ for $j\neq k$ in~(\ref{scB}). In order to calculate the scalar products
for other arrangements of the  variables one has to use symmetry properties of the eigenfunctions.
The functions $M_S(\boldsymbol{x},\boldsymbol{x}')$ take the form
\begin{equation}
M_A(\boldsymbol{x},\boldsymbol{x}')=\prod_{{j+k\leq N}}\alpha(\boldsymbol{x}_j,\boldsymbol{x}'_k)\,,
\quad
M_B(\boldsymbol{x},\boldsymbol{x}')=|p p'|^{N-1}\prod_{j+k\leq N-1}\alpha(\boldsymbol{x}_j,\boldsymbol{x}'_k)\,.
\end{equation}
The calculation of the product   $\tilde \Lambda_1^\dagger \tilde \Lambda_1$ is straightforward
\begin{equation}\fl
\tilde \Lambda_1^\dagger(\boldsymbol{x}') \tilde \Lambda_1(\boldsymbol{x})=\int d^2z {z^{-1+i(x-x')}\bar z^{-1+i(\bar x-\bar x')}}=
2\pi^2\,\delta_{nn'}\,\delta(\nu-\nu')=2\pi^2\,\delta(\boldsymbol{x}-\boldsymbol{x}')\,.
\end{equation}
Taking into account Eqs.~(\ref{alphaxx}) and (\ref{scA})  we obtain for the measure $\boldsymbol{\mu}_A(\boldsymbol{x})$
\begin{equation}\fl
\boldsymbol{\mu}_A(\boldsymbol{x})=(2\pi)^{-N} \pi^{-N^2} \prod_{k<j}(x_k-x_j)(\bar x_k-\bar x_j)=
(2\pi)^{-N} \pi^{-N^2} \prod_{j<k}\left((\nu_k-\nu_j)^2+\frac14(n_k-n_j)^2\right)\,.
\end{equation}
Further,   Eq.~(\ref{scB}) can be simplified with the help of  the following relation
\begin{equation}\label{LL2}
\Lambda_2^\dagger(\boldsymbol{x}')\,\Lambda_2(\boldsymbol{x})e^{i(p z+\bar p\bar z)}=e^{i(p z+\bar p\bar z)}|p|^{-2}\,
2\pi^4\,\delta(\boldsymbol{x}-\boldsymbol{x}')\,.
\end{equation}
In order to verify  Eq.~(\ref{LL2}) one can  calculate a diagram which corresponds to the l.h.s of this equation.
It  can be  done easily by going over to the momentum representation (see also Ref.~\cite{Derkachov:2001yn}).
Collecting all factors one gets for the measure
\begin{eqnarray}
\boldsymbol{\mu}_B(\boldsymbol{x})&=&2(2\pi)^{-N} \pi^{-N^2} \prod_{k<j\leq N-1}(x_k-x_j)(\bar x_k-\bar x_j)
\nonumber\\
&=&
2(2\pi)^{-N} \pi^{-N^2} \prod_{j<k\leq N-1}\left((\nu_k-\nu_j)^2+\frac14(n_k-n_j)^2\right)\,.
\end{eqnarray}
We would like to stress here that  the completeness of the constructed orthonormal systems
\begin{equation}\label{completeness}
\int \mathcal{D}_S \boldsymbol{x}\,\boldsymbol{\mu}_S(\boldsymbol{x})\,\Psi_{S} ({\boldsymbol{x}}|\boldsymbol{z})\,
\left(\Psi_{S} ({\boldsymbol{x}}|\boldsymbol{z}')\right)^\dagger=\prod_{k=1}^N\delta^2(\vec{z}_k-\vec{z}'_k)
\end{equation}
deserves a separate study.
For $N=1,2$ Eq.~(\ref{completeness}) can be easily checked by elementary methods, see e.g.~\cite{Derkachov:2002tf}.
As to general $N$  we hope that the method developed in~Ref.~\cite{Kozlowski} for the quantum Toda chain
could prove useful for verifying the completeness condition (\ref{completeness}).

Closing this section we want to  mention  that the basis of eigenfunctions of the elements of the momodromy matrix
proves to be useful in applications, e.g. for studies of form
factors~\cite{Smirnov:1992vz,Niccoli:2012ci,Niccoli:2012vq,Tarasov:1991mf} and correlation functions~\cite{MS,KMT}.
The
basis of eigenfunctions of
$B-$operator plays a prominent role in analysis of the closed spin chains since it determines  so-called
Sklyanin's representation of Separated Variables~\cite{Skl91}. The applications of the SoV methods for
particular models can be found in
Refs.~\cite{Sklyanin:1984sb,Kharchev:1999bh,Kharchev:2000yj,Derkachov:2001yn,Silantyev,Niccoli:2012ci,Niccoli:2012vq}.

\section{Baxter's operators}\label{sect:Baxter}

The method of Baxter's operators~\cite{Baxter:1972hz}
provides
an alternative to the conventional Algebraic Bethe Ansatz. Let operators $\mathcal{Q}(u)$ form  a commutative  family,
$[\mathcal{Q}(u),\mathcal{Q}(v)]=0$. The operator $\mathcal{Q}(u)$ is called Baxter operator if it also commutes with
integral of motions of the model (including Hamiltonian) and satisfies a certain finite-difference equation (Baxter equation).
Provided that the analytic properties of the eigenvalues as   functions of the spectral parameter $u$ are known one can obtain them by solving
the Baxter equation. It turns out that such fundamental objects  as transfer matrices and the Hamiltonian can be expressed in terms
of  $\mathcal{Q}$ operators in a rather simple way. For the closed $SL(2,\mathbb{C})$ spin chains  the Baxter operators
were
constructed in Ref.~\cite{Derkachov:2001yn}.
In this section we construct the set of Baxter operators $\mathcal{Q}_S(\boldsymbol{u})\equiv\mathcal{Q}_S(u,\bar u)$,
\ $S=A,B,C,D$,\  such that they
commute with the corresponding elements of the monodromy matrices, ${T}_N(v)$, $\bar{{T}}_N(\bar v)$,
\begin{eqnarray}
[\mathcal{Q}_A(\boldsymbol{u}),\ A_N(v)]& = &[\mathcal{Q}_A(\boldsymbol{u}),\ \bar A_N(\bar v)]=0\,,
\nonumber\\{}
[\mathcal{Q}_D(\boldsymbol{u}),\ D_N(v)]& = &[\mathcal{Q}_D(\boldsymbol{u}),\ \bar D_N(\bar v)]=0\,,
\end{eqnarray}
etc. We also derive   the difference equations which these operators satisfy.

\begin{figure}[t]
\centerline{\includegraphics[width=0.98\linewidth]{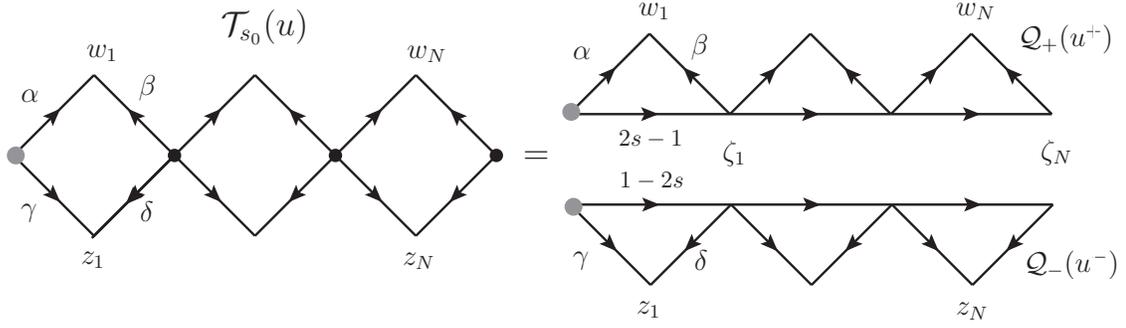}}
\caption{The diagrammatic representation of the operator $\mathcal{T}_{\boldsymbol{s}_0}(\boldsymbol{u})=\mathcal{Q}_-(u+i(1-s_0))
\mathcal{Q}_+(u+is_0)$.
Here the black blobs stand for the integration vertices, the gray blobs indicate $z_0=0$.
The indices are the following $\alpha=1-s-iu^+$, $\beta=1-s+iu^+$, $\gamma=s+iu^-$,  $\delta=s-iu^-$
and $u^-=u+i(1-s_0)$, $u^+=u+is_0$
}
\label{OmegaQQ}
\end{figure}

Let us define an operator
\begin{equation}
\mathbb{T}_{\boldsymbol{s}_0}(\boldsymbol{u})=\mathcal{R}_{\boldsymbol{s}_0 \boldsymbol{s}_1}(\boldsymbol{u})\,
\mathcal{R}_{\boldsymbol{s}_0 \boldsymbol{s}_{2}}(\boldsymbol{u})
\ldots\mathcal{R}_{\boldsymbol{s}_0 \boldsymbol{s}_N} (\boldsymbol{u})\,
\end{equation}
 which acts on the tensor product $\mathbb{V}_0\otimes
\mathbb{H}_N$, where $\mathbb{V}_0=L_2(\mathbb{C})$ is an auxiliary space
and $\mathbb{H}_N$ is the quantum space of the model, $\mathbb{H}_N=\otimes_{k=1}^N \mathbb{V}_k$.
As usual it is assumed that the operator $\mathcal{R}_{\boldsymbol{s}_0 \boldsymbol{s}_k}(\boldsymbol{u})$, see Eq.~(\ref{RLL}),
acts nontrivially on the tensor product
$\mathbb{V}_0\otimes\mathbb{V}_k$.
It follows from the  relation~(\ref{RLL}) that this operator obeys the following commutation relation
\begin{equation}\label{TLL}
\mathbb{T}_{\boldsymbol{s}_0}(\boldsymbol{u})\, {T}_N(v)\,L_{s_0} (v-u)=
 L_{s_0} (v-u)\, {T}_N(v)\,\mathbb{T}_{\boldsymbol{s}_0}(\boldsymbol{u})\,,
\end{equation}
where ${T}_N(v)$ is the monodromy matrix~(\ref{monodromy}) and $L_{s_0}$ is the $L-$operator which acts on
$\mathbb{V}_0\otimes \mathbb{C}^2$. The  products ${T}_N(v)\,L_{s_0} (v-u)$ and
$ L_{s_0} (v-u)\, {T}_N(v)\,$ are $2\times 2$ matrices so that  Eq.~(\ref{TLL}) reads in explicit form
\begin{eqnarray}\fl
\mathbb{T}_{\boldsymbol{s}_0}(\boldsymbol{u})
\left(\begin{array}{cc}
A_N(v) & B_N(v)\\
C_N(v) & D_N(v)
\end{array}\right)
\left(\begin{array}{cc}
v-u+z_0\partial_{z_0}+s_0 & -\partial_{z_0}\\
z_0^2\partial_{z_0}+2s_0z_0 & v-u-z_0\partial_{z_0}-s_0
\end{array}\right)
=
\nonumber\\
=
\left(\begin{array}{cc}
v-u+z_0\partial_{z_0}+s_0 & -\partial_{z_0}\\
z_0^2\partial_{z_0}+2s_0z_0 & v-u-z_0\partial_{z_0}-s_0
\end{array}\right)
\left(\begin{array}{cc}
A_N(v) & B_N(v)\\
C_N(v) & D_N(v)
\end{array}\right)
\mathbb{T}_{\boldsymbol{s}_0}(\boldsymbol{u})\,.
\end{eqnarray}
The equation involving the matrix element $(2,2)$ has the form
\begin{eqnarray}\label{RD}
\fl\mathbb{T}_{\boldsymbol{s}_0}(\boldsymbol{u})\Big(D_N(v)(v-u-z_0\partial_{z_0}-s_0) - C_{N}(v)\partial_{z_0} \Big)
=
\nonumber\\
=\Big((v-u-z_0\partial_{z_0}-s_0) D_N(v)+
(z_0^2\partial_{z_0}+2s_0z_0)B_N(v)\Big)\mathbb{T}_{\boldsymbol{s}_0}(\boldsymbol{u}).
\end{eqnarray}
The l.h.s and r.h.s. of this equation are  operators that act on the space of functions of $N+1$ variables,
$\psi(z_0,z_1,\ldots,z_N)$.
Applying both sides of Eq.~(\ref{RD}) to the function $f=f(z_1,\ldots,z_N)$ which does not depend on
$z_0$ and sending $z_0\to 0$ in the result one obtains
\begin{equation}
\mathbb{T}_{\boldsymbol{s}_0}(\boldsymbol{u})\,D_N(v) f=D_N(v)\,\mathbb{T}_{\boldsymbol{s}_0}(\boldsymbol{u})f\Big|_{z_0=0}\,.
\end{equation}
{Hence the operator $\mathcal{T}_{\boldsymbol{s}_0}(\boldsymbol{u})$ which is defined on the space of functions of $N$
variables }
\begin{equation}
[\mathcal{T}_{\boldsymbol{s}_0}(\boldsymbol{u}) f](z_1,\ldots,z_N)=[\mathbb{T}_{\boldsymbol{s}_0}(\boldsymbol{u})f](z_0,z_1,\ldots,z_N)|_{z_0\to 0}
\end{equation}
commutes with the element  $D_N(v)$ (and $\bar D_N(\bar v)$) of the monodromy matrix
\begin{equation}
[\mathcal{T}_{\boldsymbol{s}_0}(\boldsymbol{u}), D_N(v)]=[\mathcal{T}_{\boldsymbol{s}_0}(\boldsymbol{u}), \bar D_N(\bar v)]=0\,.
\end{equation}
The kernel of the integral operator $\mathcal{T}_{\boldsymbol{s}_0}(\boldsymbol{u})$
is related to the kernel of the operator $\mathbb{T}_{\boldsymbol{s}_0}(\boldsymbol{u})$ as follows
\begin{equation}
\mathcal{T}_{\boldsymbol{s}_0}(\boldsymbol{u})(z_1,\ldots,z_N|w_1,\ldots,w_N)=
\int d^2 w\, \mathbb{T}_{\boldsymbol{s}_0}(\boldsymbol{u})(0,z_1,\ldots,z_N|w,w_1,\ldots,w_N)\,.
\end{equation}
The operator $\mathcal{T}_{\boldsymbol{s}_0}(\boldsymbol{u})$ depends on the spins $s_0, \bar s_0$ in the auxiliary space
and the spectral parameters $u,\bar u$
and thus can be considered as an analog of a transfer matrix.
{The proof of commutativity
\begin{equation}\label{Om2}
[\mathcal{T}_{\boldsymbol{s}_0}(\boldsymbol{u}),\mathcal{T}_{\boldsymbol{s}_{0'}}(\boldsymbol{v})]=0\,.
\end{equation}
is given in Appendix~\ref{App:B}.}

It is known that the transfer matrices for the
$SL(2,\mathbb{C})$
spin chains  factorize into the
product of two Baxter $\mathcal{Q}$ operators~\cite{Derkachov:2001yn}. The same holds true in the case under
consideration.
The operator $\mathcal{T}_{\boldsymbol{s}_0}(\boldsymbol{u})$ can be represented as product of two operators
\begin{equation}
\mathcal{T}_{\boldsymbol{s}_0}(\boldsymbol{u})=\mathcal{Q}_-(\boldsymbol{u}+i(1-\boldsymbol{s}_0))\,
\mathcal{Q}_+(\boldsymbol{u}+i\boldsymbol{s}_0).
\end{equation}
The kernel of  $\mathcal{T}_{\boldsymbol{s}_0}(\boldsymbol{u})$ and its representation in the factorized form
is shown in Fig.~\ref{OmegaQQ}. While the ``transfer matrix'' $\mathcal{T}_{\boldsymbol{s}_0}(\boldsymbol{u})$ depends on two
sets of variables: the spin $\boldsymbol{s}_0=(s_0,\bar s_0)$ and the spectral parameter $\boldsymbol{u}=(u,\bar u)$, each
of
the operators $\mathcal{Q}_\pm$ depends only on one variable. In the explicit form the kernels of the operators
$\mathcal{Q}_\pm(\boldsymbol{u}))$ are given by the following  expressions
\begin{eqnarray}\fl
\mathcal{Q}_-(\boldsymbol{u})(z_1,\ldots, z_N|w_1,\ldots, w_N)& = &
\prod_{k=1}^N[z_k-w_{k-1}]^{-s-iu}[z_k-w_{k}]^{-s+iu}[w_{k}-w_{k-1}]^{2s-1}\,,
\nonumber\\
\fl
\mathcal{Q}_+(\boldsymbol{u})(z_1,\ldots, z_N|w_1,\ldots, w_N)& = &
\prod_{k=1}^N[w_{k}-z_{k-1}]^{-1+s+iu}[w_{k}-z_k]^{-1+s-iu}[z_{k}-z_{k-1}]^{1-2s}\,,
\end{eqnarray}
where $w_0=z_0=0$. The requirement for the kernel to be a single-valued  function on the complex plane
results in the following restriction on the spectral parameters
\begin{equation}
(s-\bar s)+i(u-\bar u)=n_s+i(u-\bar u)=n\in \mathbb{Z}\,.
\end{equation}
Thus the spectral parameters $u,\bar u$ have the form~(\ref{x-n-nu}) where $\nu$ takes now complex values.

Taking into account Eq.~(\ref{delta1}) one easily derives that operators $\mathcal{Q}_\pm$ satisfy the following
normalization conditions
\begin{equation}\label{Qpm-normalization}\fl
\mathcal{Q}_-(i(1-\boldsymbol{s})+\epsilon)=\left(\frac{\pi}{i\epsilon}\right)^N \Big(\II+O(\epsilon)\Big)\,,
\qquad
\mathcal{Q}_+(-i\boldsymbol{s}-\epsilon)=\left(\frac{\pi}{i\epsilon}\right)^N \Big(\II+O(\epsilon)\Big)\,.
\end{equation}
Eqs.~(\ref{Qpm-normalization}) allow one  to represent the  operators $\mathcal{Q}_\pm$ as  certain limits
of the operator
$\mathcal{T}_{\boldsymbol{s}_0}(\boldsymbol{u})$, e.g.
\begin{equation}
\lim_{\epsilon\to 0}(i\epsilon)^N\mathcal{T}_{i\boldsymbol{u}-\boldsymbol{s}+i\epsilon}(\boldsymbol{u})=\pi^N
\mathcal{Q}_-(2\boldsymbol{u}+i(1-\boldsymbol{s}))\,.
\end{equation}
This implies that each of  the Baxter operators $\mathcal{Q}_\pm(\boldsymbol{u})$ commute with the operator $D_N(u)$
\begin{equation}
[\mathcal{Q}_\pm(\boldsymbol{u}), D_N(u)]=[\mathcal{Q}_\pm(u), \bar D_N(\bar u)]=0\,.
\end{equation}
The commutativity of the operators $\mathcal{Q}_\pm(\boldsymbol{u})$
\begin{equation}
[\mathcal{Q}_+(\boldsymbol{u}),\mathcal{Q}_+(\boldsymbol{v})]=[\mathcal{Q}_-(\boldsymbol{u}),\mathcal{Q}_-(\boldsymbol{v})]
=[\mathcal{Q}_+(\boldsymbol{u}),\mathcal{Q}_-(\boldsymbol{v})]=0,
\end{equation}
can be checked diagrammatically with the help of the identities given in Appendix~\ref{Diagram}. Alternatively,  it can
be derived from the commutativity of the operators $\mathcal{T}_{\boldsymbol{s}_0}(\boldsymbol{u})$.
Since the operators $\mathcal{Q}_\pm$ are related by the hermitian conjugation,
$\mathcal{Q}_+(u,\bar u)=(\mathcal{Q}_-(\bar u^*, u^*))^\dagger$,
it is sufficient to consider only one of them.  Let
\begin{equation}
\mathcal{Q}_D(\boldsymbol{u})\equiv\mathcal{Q}_-(\boldsymbol{u}).
\end{equation}
The operator $\mathcal{Q}_D(\boldsymbol{u})$ satisfy the finite-difference equations:
\begin{eqnarray}\label{D-dif}
D_N(u)\mathcal{Q}_D(u,\bar u)&=&(u+is)^N \mathcal{Q}_D(u+i,\bar u)\,,
\nonumber\\
\bar D_N(\bar u)\mathcal{Q}_D(u,\bar u)&=&(\bar u+i\bar s)^N \mathcal{Q}_D(u,\bar u+i)\,.
\end{eqnarray}
These equations can be derived making use of the invariance of the
the monodromy matrices under ``gauge'' rotations of
$L-$operators: $L_k\to M_{k-1}L_k M_k$, with
$M_k=\left(\begin{array}{cc}
1&0\\ -w_k& 1
\end{array}\right)$,
with $w_0=0$. We will not dwell on this derivation here since
this method was discussed in great detail in~\cite{Derkachov:1999pz,Derkachov:2001yn,Derkachov:2002tf}.

To summarize, we have constructed the commutative family of the operators $\mathcal{Q}_D(\boldsymbol{u})$
with the following properties:
\begin{itemize}
\item $[\mathcal{Q}_D(\boldsymbol{u}),\mathcal{Q}_D(\boldsymbol{v})]=0$
\item $[D_N(u),\mathcal{Q}_D(\boldsymbol{v})]=[\bar D_N(\bar u),\mathcal{Q}_D(\boldsymbol{v})]=0$
\item $\mathcal{Q}_D(\boldsymbol{u})$ satisfy the difference - equations~(\ref{D-dif})
\item $\mathcal{Q}_D(i(1-\boldsymbol{s})+\epsilon)=\left(\frac{\pi}{i\epsilon}\right)^N \Big(\II+O(\epsilon)\Big)$
\end{itemize}
The operators $\mathcal{Q}_D(\boldsymbol{u})$ and $D_N(u),\,\bar D_N(\bar u)$ share the same eigenfunctions.
The eigenfunctions of the operators $D_N(u),\,\bar D_N(\bar u)$, $\Psi_D(\boldsymbol{x}|\boldsymbol{z})$,
were constructed in Sect.~\ref{sect:iterative}, Eq.~(\ref{PsiD}).
Thus we conclude that
\begin{equation}
\mathcal{Q}_D(\boldsymbol{u})\Psi_D(\boldsymbol{x}|\boldsymbol{z})=
q_D(\boldsymbol{u},\boldsymbol{x})\Psi_D(\boldsymbol{x}|\boldsymbol{z})\,.
\end{equation}
The eigenvalue $q_D$ is given by the following expression
\begin{equation}\label{QD-eigenvalue}
q_D(\boldsymbol{u},\boldsymbol{x})=\pi^N\prod_{k=1}^Na(1+i\bar u-i\bar x_k, s-i u,1-s+ix_k)\,,
\end{equation}
which can  be easily found with the help of the following identity
\begin{equation}
\mathcal{Q}^{(N)}_D(\boldsymbol{u})\,\widehat \Lambda_N(\boldsymbol{x})=
\pi\, a(1+i\bar u-i\bar x, s-i u,1-s+i  x)\,
\widehat \Lambda_N(\boldsymbol{x})\,\mathcal{Q}^{(N-1)}_D(\boldsymbol{u})\,.
\end{equation}
Proceeding along the same lines one can construct Baxter operators for all other cases. We will skip details and present only
final expressions for the kernels, difference equations and normalization of the Baxter operators.
\begin{itemize}
\item
$\mathcal{Q}_A(\boldsymbol{u})$ operator:
\begin{description}
\item{i. Kernel} (below $\boldsymbol{z}=(z_1,\ldots,z_N)$, $\boldsymbol{w}=(w_1,\ldots,w_N)$,  $w_0=w_{N+1}=0$)
\begin{equation}\fl
\mathcal{Q}_A(\boldsymbol{u})(\boldsymbol{z}|\boldsymbol{w})=
\prod_{k=1}^N[z_k-w_{k}]^{-s-iu}[z_k-w_{k+1}]^{-s+iu}[w_{k}-w_{k+1}]^{2s-1}\,,
\end{equation}
\item{ii. Difference equations}
\begin{equation*}\fl
A_N(u)\mathcal{Q}_A(u,\bar u)=(u-is)^N \, \mathcal{Q}_A(u-i,\bar u)\,,
\quad
\bar A_N(\bar u)\mathcal{Q}_A(u,\bar u)=(\bar u-i\bar s)^N \, \mathcal{Q}_A(u,\bar u-i)\,.
\end{equation*}
\item{iii. Normalization}
\begin{equation*}\fl
\mathcal{Q}_A(-i(1-\boldsymbol{s})-\epsilon)=\left(\frac{\pi}{i\epsilon}\right)^N\,\Big(\II+O(\epsilon)\Big)\,.
\end{equation*}
\end{description}
\item $\mathcal{Q}_B(\boldsymbol{u})$ operator:
\begin{description}
\item{i. Kernel}
\begin{equation}\fl
\mathcal{Q}_B(\boldsymbol{u})(\boldsymbol{z}|\boldsymbol{w})=[z_1-w_1]^{-s+iu}
\prod_{k=2}^N[z_k-w_{k-1}]^{-s-iu}[z_k-w_{k}]^{-s+iu}[w_{k}-w_{k-1}]^{2s-1}\,,
\end{equation}
\item{ii. Difference equations}
\begin{equation*}\fl
B_N(u)\mathcal{Q}_B(u,\bar u)=(u+is)^N \, \mathcal{Q}_B(u+i,\bar u)\,,
\quad
\bar B_N(\bar u)\mathcal{Q}_B(u,\bar u)=(\bar u+i\bar s)^N \, \mathcal{Q}_B(u,\bar u+i)\,.
\end{equation*}
\item{iii. Normalization}
\begin{equation*}\fl
\mathcal{Q}_B(i(1-\boldsymbol{s})+\epsilon)=\left(\frac{\pi}{i\epsilon}\right)^N\,\Big(\II+O(\epsilon)\Big)\,.
\end{equation*}
\end{description}

\item $\mathcal{Q}_C(\boldsymbol{u})$ operator:

\begin{description}
\item{i. Kernel}
\begin{equation}\fl
\mathcal{Q}_C(\boldsymbol{u})(\boldsymbol{z}|\boldsymbol{w})=[-w_N]^{-1+s-iu}
\prod_{k=1}^N[z_k-w_{k-1}]^{-s-iu}[z_k-w_{k}]^{-s+iu}[w_{k}-w_{k-1}]^{2s-1}\,,
\end{equation}
\item{ii. Difference equations}
\begin{equation*}\fl
C_N(u)\mathcal{Q}_C(u,\bar u)=(u+is)^N \, \mathcal{Q}_A(u+i,\bar u)\,,
\quad
\bar C_N(\bar u)\mathcal{Q}_C(u,\bar u)=(\bar u+i\bar s)^N \, \mathcal{Q}_A(u,\bar u+i)\,.
\end{equation*}
\item{iii. Normalization}
\begin{equation*}\fl
\mathcal{Q}_C(-i(1-\boldsymbol{s})-\epsilon)=\left(\frac{\pi}{i\epsilon}\right)^N\,\Big(\II+O(\epsilon)\Big)\,.
\end{equation*}
\end{description}

\item $\mathcal{Q}_D(\boldsymbol{u})$ operator:
\begin{description}
\item{i. Kernel}
\begin{equation}\label{QD-kernel}\fl
\mathcal{Q}_D(\boldsymbol{z}|\boldsymbol{w})=
\prod_{k=1}^N[z_k-w_{k-1}]^{-s-iu}[z_k-w_{k}]^{-s+iu}[w_{k}-w_{k-1}]^{2s-1}\,,
\end{equation}
\item{ii. Difference equations}
\begin{equation*}\fl
D_N(u)\mathcal{Q}_B(u,\bar u)=(u+is)^N \, \mathcal{Q}_D(u+i,\bar u)\,,
\quad
\bar D_N(\bar u)\mathcal{Q}_D(u,\bar u)=(\bar u+i\bar s)^N \, \mathcal{Q}_D(u,\bar u+i)\,.
\end{equation*}
\item{iii. Normalization}
\begin{equation*}\fl
\mathcal{Q}_D(i(1-\boldsymbol{s})+\epsilon)=\left(\frac{\pi}{i\epsilon}\right)^N\,\Big(\II+O(\epsilon)\Big)\,.
\end{equation*}
\end{description}
\end{itemize}

%
%

\section{Hamiltonians}\label{sect:Hamiltonian}
One can generate integrable  Hamiltonians
calculating further terms in the $\epsilon$~-~expansion of the Baxter operators at the special points,
$\boldsymbol{u}=\pm (i(1-\boldsymbol{s})+\epsilon)$. We will construct the Hamiltonian which commutes with
the elements of the transfer matrices $D_N(u)$, $\bar
D_N(\bar u)$. This Hamiltonian has appeared   in the studies of the scattering
amplitudes in the Regge limit in the $\mathcal{N}=4$ SUSY~\cite{Lipatov:2009nt,Bartels:2011nz}.

To work out the $\epsilon$ expansion of the operator $\mathcal{Q}_D$ it is convenient to use the
equivalent representation
\begin{equation}\label{Q-D2}
\mathcal{Q}_D(-i(1-\boldsymbol{s})+\boldsymbol{\epsilon})=
\left. \mathcal{R}_{01}(\boldsymbol{\epsilon})
\mathcal{R}_{12}(\boldsymbol{\epsilon})\ldots
\mathcal{R}_{N-1,N}(\boldsymbol{\epsilon})
\right|_{z_0 = 0} \,,
\end{equation}
where the operator $\mathcal{R}_{kk+1}(\boldsymbol{\epsilon})$ is defined as follows
\begin{eqnarray}\label{Rkk}\fl
[\mathcal{R}_{kk+1}(\boldsymbol{\epsilon}) f](\ldots,z_{k},z_{k+1},\ldots)=
\nonumber\\{}
=\int d^2 w_{k+1}
\frac{[w_{k+1}-z_{k}]^{2s-1}}{[z_{k+1}-z_{k}]^{2s-1+i\epsilon}[z_{k+1}-w_{k+1}]^{1-i\epsilon}} f(\ldots,z_{k},w_{k+1},\ldots).
\end{eqnarray}
Making use of  Eq.~(\ref{Rkk}) it is straightforward to verify that the kernel of the operator
$\mathcal{Q}_D(-i(1-\boldsymbol{s})+\boldsymbol{\epsilon})$ in Eq.~(\ref{Q-D2})  has the  form~(\ref{QD-kernel}).
Note also that the operator
$\mathcal{R}_{kk+1}$ is nothing else as the factorizing operator $\mathcal{R}^{(1)}_{kk+1}$ for the
special choice of  spectral parameters, see Eq.~(\ref{Rint}).
This is a unitary operator provided that $\epsilon=\bar\epsilon^*$ (that will be implied henceforth)
\begin{equation}
(\mathcal{R}_{kk+1}(\boldsymbol{\epsilon}))^\dagger\mathcal{R}_{kk+1}(\boldsymbol{\epsilon})=
\left(\frac{\pi}{\epsilon}\right)^2\II\,.
\end{equation}
The operator  $\mathcal{R}_{kk+1}(\boldsymbol{\epsilon})$ can be represented in several different forms
\begin{eqnarray}\label{twoform}
\mathcal{R}_{kk+1}(\boldsymbol{\epsilon})&=&\pi\, a(1-i\epsilon)\,
[z_{kk+1}]^{1-2s-i\epsilon}\,[i\partial_{k+1}]^{-i\epsilon}\, [z_{kk+1}]^{2s-1}
\nonumber\\[2mm]
&=&\pi\, a(1-i\epsilon)\,
[i\partial_{k+1}]^{2s-1}\, [z_{kk+1}]^{-i\epsilon}\,[i\partial_{k+1}]^{1-2s-i\epsilon}
\nonumber\\
&=&\pi\, a(1-i\epsilon)\frac{\Gamma(2\bar s-\bar z_{kk+1}\bar\partial_{k+1})}{\Gamma(2\bar s-\bar z_{kk+1}\bar\partial_{k+1}+i\epsilon)}
\frac{\Gamma(1-2s+z_{kk+1}\partial_{k+1}-i\epsilon)}{\Gamma(1-2s+z_{kk+1}\partial_{k+1})}\,.
\end{eqnarray}
Here $z_{kk+1}=z_k-z_{k+1}$ and $\partial_k=\partial_{z_k}, \, \bar \partial_k=\partial_{\bar z_k}$. The
operator
$[i\partial_z]^{\alpha}$ is defined as
an operator of multiplication by $[p]^\alpha$ in the momentum space, i.e.
\begin{equation}
[i\partial_z]^{\alpha} \,f(z)=[i\partial_z]^{\alpha} \,\int d^2 p \,f(p)\,e^{-i(pz+\bar p\bar z)}=\int d^2 p \,
[p]^\alpha\,f(p)\,e^{-i(pz+\bar p\bar z)}\,.
\end{equation}
The first line of Eq.~(\ref{twoform})  follows directly from~ Eq.~(\ref{Rkk}) and Eq.~(\ref{FD}).
It can be cast into the form given in the second line with the help of Eq.~(\ref{STOP}).
Further, let us represent the operator in the first line  as follows
\begin{eqnarray*}\fl
[z_{kk+1}]^{1-2s-i\epsilon}\,[i\partial_{k+1}]^{-i\epsilon}\, [z_{kk+1}]^{2s-1}=
\nonumber\\
e^{-z_{k}\partial_{k+1}-\bar z_{k}\bar\partial_{k+1}}\,
\Big([z_{k+1}]^{1-2s-i\epsilon}[i\partial_{k+1}]^{-i\epsilon} [z_{k+1}]^{2s-1} \Big) \,e^{z_{k}\partial_{k+1}
+\bar z_{k}\bar\partial_{k+1}}\,.
\end{eqnarray*}
Obviously, the operator in the  brackets  (we change $z_{k+1}\to z$)
\begin{equation}
F=[z]^{1-2s-i\epsilon}[i\partial]^{-i\epsilon} [z]^{2s-1}
\end{equation}
commutes with the operators $z\partial_z$ and $\bar z\partial_{\bar z}$. Therefore the  power
$f_\alpha(z)=[z]^\alpha=z^\alpha\bar z^{\bar \alpha}$ are  eigenfunctions of $F$: $[F
f_\alpha](z)=\mathcal{F}(\alpha,\bar\alpha)f_\alpha[z]$, where $\mathcal{F}(\alpha,\bar\alpha)$ is the
corresponding eigenvalue. As a consequence we can represent the operator $F$ in the following form
$F=\mathcal{F}(z\partial_z,\bar z\partial_{\bar z})$.
Calculating  the eigenvalue $\mathcal{F}(\alpha,\bar\alpha)$ with the help of
Eqs.~(\ref{FD}),~(\ref{Chain}) one obtains the representation for the operator
$\mathcal{R}_{kk+1}(\boldsymbol{\epsilon})$ given in  the third line of Eq.~(\ref{twoform}).

Making use of  Eq.~(\ref{twoform}) one can easily  find first terms in the $\epsilon$ expansion of the
operator
$\mathcal{R}_{kk+1}(\boldsymbol{\epsilon})$
\begin{equation}
\mathcal{R}_{kk+1}(\boldsymbol{\epsilon})=\frac{\pi}{i\epsilon}
\Big(\II-i\epsilon\,\mathcal{H}_{kk+1}+O(\epsilon^2)\Big)\,,
\end{equation}
where the pair-wise Hamiltonian $\mathcal{H}_{k-1k}$ reads~\footnote{Formally, the Hamiltonian
in Eq.~(\ref{Hkk}) splits up  in the sum of two operators acting in the holomorphic and anti-holomorphic sectors,
respectively.
We want to stress here that  these two operators have to be considered separately with certain care since
only their sum  presents a well-defined object.}
\begin{eqnarray}\label{Hkk}
\mathcal{H}_{kk+1}&=&\ln[z_{kk+1}]+[z_{kk+1}]^{1-2s} \ln[i\partial_{k+1}]\,[z_{kk+1}]^{2s-1}-2\psi(1)
\nonumber\\[1mm]
&=&
\ln[i\partial_{k+1}]+[i\partial_{k+1}]^{2s-1}\,\ln[z_{kk+1}]\,[i\partial_{k+1}]^{1-2s}-2\psi(1)
\nonumber\\[1mm]
&=&
\psi(1-2s+z_{kk+1}\partial_{k+1})+\psi(2\bar s-\bar z_{kk+1}\bar\partial_{k+1})-2\psi(1)
\nonumber\\
&=&\psi(2s-z_{kk+1}\partial_{k+1})+\psi(1-2\bar s+\bar z_{kk+1}\bar\partial_{k+1})-2\psi(1)\,.
\end{eqnarray}
Here $\psi(x)=(\log \Gamma(x))'$ is the Euler $\psi$ function, $\ln[z_{kk+1}]=\ln (z_{kk+1} \bar
z_{kk+1})=2\ln|z_{kk+1}|$ and $\ln[i\partial_k]=\ln(-\partial_k\bar \partial_k)$.

The pair-wise Hamiltonians
$\mathcal{H}_{kk+1}$
are evidently  self-adjoint operators, $\mathcal{H}_{kk+1}=\mathcal{H}_{kk+1}^\dagger$.
Note that the Hamiltonian $\mathcal{H}_{kk+1}$ is not  $SL(2,\mathbb{C})$ invariant operator.
It commutes with two of three the  $SL(2,\mathbb{C})$  generators (we discuss the holomorphic sector only)
\begin{eqnarray*}
S^{(+)}_{kk+1}&=&z_{k+1}^2\partial_{k+1}+z_k^2\partial_k+2s(z_{k}+z_{k+1})\,,  \qquad S^{(-)}_{kk+1}=-\partial_{k}-\partial_{k+1}\,,
\nonumber\\
 S^{(0)}_{kk+1}&=&z_{k}\partial_{k}+z_{k+1}\partial_{k+1}+2s\,.
\end{eqnarray*}
Namely,
\begin{equation}
[S^{(-)}_{kk+1},\mathcal{H}_{kk+1}]=[S^{(0)}_{kk+1},\mathcal{H}_{kk+1}]=0
\end{equation}
whereas
\begin{equation}
[S^{(+)}_{kk+1},\mathcal{H}_{kk+1}]=z_{k}-z_{k+1}.
\end{equation}
To derive the last equation it is sufficient to notice that the operator $\mathcal{R}_{kk+1}(\boldsymbol{\epsilon})$
intertwines the tensor products of the $SL(2,\mathbb{C})$ representations
\begin{equation}\label{RTT}
\mathcal{R}_{kk+1}(\boldsymbol{\epsilon})\,T^{\boldsymbol{s}}\otimes T^{\boldsymbol{s}+i\epsilon/2}=
T^{\boldsymbol{s}+i\epsilon/2}\otimes T^{\boldsymbol{s}}\,\mathcal{R}_{kk+1}(\boldsymbol{\epsilon})\,.
\end{equation}
It can be easily checked with the help of Eq.~(\ref{Tg}).  In  turn Eq.~(\ref{RTT})  implies
\begin{equation}
\mathcal{R}_{kk+1}(\boldsymbol{\epsilon})\,(S^{(+)}_{kk+1}+i\epsilon z_{k+1})=
(S^{(+)}_{kk+1}+i\epsilon z_{k})\,\mathcal{R}_{kk+1}(\boldsymbol{\epsilon})\,.
\end{equation}

Collecting everything we obtain
\begin{equation}\label{QDexp}
\mathcal{Q}_D(-i(1-\boldsymbol{s})+\boldsymbol{\epsilon})=\left(\frac{\pi}{i\epsilon}\right)^N
\Big(\II-i\epsilon \,\mathcal{H}_N+O(\epsilon^2)\Big)\,,
\end{equation}
where
\begin{equation}
\mathcal{H}_N=\sum_{k=0}^{N-1}\mathcal{H}_{kk+1}
\end{equation}
is a self-adjoint operator $\mathcal{H}_N=\mathcal{H}_N^\dagger$. We stress here that the pair-wise Hamiltonians are not
$SL(2,\mathbb{C})$ invariant.~\footnote{Let us note that in the case of the closed $SL(2,\mathbb{C})$ magnet the situation is
exactly the same. The Hamiltonians given by the derivative of Baxter operator at the point $u=\pm i(1-s)$,
$\mathcal{H}_N^\pm=(\ln Q(\pm i(1-s)))'$
 are self-adjoint
and $SL(2,\mathbb{C})$ invariant operators. Each of the Hamiltomnians $\mathcal{H}^\pm_N$ is
 given by the sum of pair Hamiltonians which are  self-adjoint
but not $SL(2,\mathbb{C})$ invariant.  However the sum of the operators,
$\mathcal{H}_N^++\mathcal{H}_N^-$,
can be represented in the form $\sum_k \mathcal{H}_{kk+1}$ where pair operators is explicitly $SL(2,\mathbb{C})$
invariant.}

\subsection{Twin Hamiltonian}

In the case of the $SL(2,\mathbb{C})$ spin chains there exists a simple method for the construction of  new operators
in  commutative  families. For definiteness we will consider
the $D$-family. The method is based on the equivalence of the $SL(2,\mathbb{C})$
representations  $T^{\boldsymbol{s}}$ and $T^{1-\boldsymbol{s}}$~\cite{Gelfand}. These representations are intertwined by the
operator $[i\partial]^{1-2s}$
\begin{equation}\label{s1ms}
[i\partial]^{1-2s}\,T^{\boldsymbol{s}}=T^{1-\boldsymbol{s}}\,[i\partial]^{1-2s}\,.
\end{equation}
Let us consider two spin chain models with  the spins $\boldsymbol{s}$ and $1-\boldsymbol{s}$, respectively.
It is natural to expect that the operators in the commutative families in these two models are related to each other.
Indeed, the elements of the monodromy matrices $D_N^{(s)}$ and  $\bar D_N^{(\bar s)}$
are linear functions of the generators, $S_{\pm,0}^{(k)}$, $k=1,\ldots,N$. Taking into account that the operator
$[i\partial]^{1-2s}$ intertwines the generators with spin $s$ and $1-s$ one immediately gets
\begin{equation} \label{WtW}
D_N^{(s) }(u)=W_N \, D_N^{(1-s)}(u) W_N^\dagger\,,
\end{equation}
where the unitary operator $W_N$ has the form
\begin{equation}
W_N=[i\partial_1]^{2s-1}\ldots [i\partial_N]^{2s-1}\,.
\end{equation}

Let us  consider an operator $\mathcal{O}^{(\boldsymbol{s})}$ from the commutative family of the first model
and its twin, $\mathcal{O}^{(1-\boldsymbol{s})}$, from the second model, i.e.
\begin{equation*}\fl
[\mathcal{O}^{(\boldsymbol{s})}, D_N^{(s)}(v)]=[\mathcal{O}^{(\boldsymbol{s})}, \bar D_N^{(\bar s)}(\bar v)]=0\,,
\quad
[\mathcal{O}^{(1-\boldsymbol{s})}, D_N^{(1-s)}(v)]=[\mathcal{O}^{(1-\boldsymbol{s})}, \bar D_N^{(1-\bar s)}(\bar v)]=0\,.
\end{equation*}
Evidently, $\widetilde{\mathcal{O}}^{(\boldsymbol{s})}= W_N
\,\mathcal{O}^{(1-\boldsymbol{s})} W_N^\dagger$ commutes with $D_N^{(s)}(v)$ and $\bar D_N^{(\bar s)}(\bar
v)$, i.e. it belongs to the first family. Moreover, in the general case  when
$\mathcal{O}^{(\boldsymbol{s})}$ is not solely a function of the spin generators~$S_k$,
the operators $\mathcal{O}^{(\boldsymbol{s})}$ and
$\widetilde{\mathcal{O}}^{(\boldsymbol{s})}$ do not necessarily coincide.
The transformation $\mathcal{O}^{(\boldsymbol{s})}\mapsto\widetilde{\mathcal{O}}^{(\boldsymbol{s})}$,
proves to be very useful and allows one to construct  new operators with required properties.
We apply it below for constructing of the new Hamiltonian.

Using the representation for   $\mathcal{H}_{kk+1}$ given in the second line
Eq.~(\ref{Hkk}) one easily finds
\begin{equation}
 W_N\, \mathcal{H}_{kk+1}^{(1-s)}\, W_N^\dagger=
\ln[i\partial_{k+1}]+[i\partial_{k}]^{2s-1}\,\ln[z_{kk+1}]\,[i\partial_{k}]^{1-2s}-2\psi(1)
\end{equation}
for $k=1,\ldots, N-1$ while for $k=0$ one gets
\begin{equation}
 W_N\, \mathcal{H}_{01}^{(1-s)}\, W_N^\dagger=
\ln [i\partial_1]+\ln[z_{1}]-2\psi(1).
\end{equation}
Writing down the expression for
$\widetilde{\mathcal{H}}_N= W_N^\dagger\, \mathcal{H}_{N}^{(1-s)}\, W_N^\dagger$ it is useful to make
some regrouping and represent the result in the following form
\begin{equation}
\widetilde{\mathcal{H}}_N=\ln[z_1]+\sum_{k=1}^{N-1}\widetilde{\mathcal{H}}_{kk+1}+\ln [i\partial_N]-2\psi(1)\,,
\end{equation}
where
\begin{eqnarray}
\widetilde{\mathcal{H}}_{kk+1}&=&\ln [i\partial_k]+[i\partial_k]^{2s-1}\ln[z_{kk+1}][i\partial_k]^{1-2s}-2\psi(1)
\nonumber\\[1mm]
&=&
\ln[z_{kk+1}]+[z_{kk+1}]^{1-2s}\ln [i\partial_k][z_{kk+1}]^{2s-1}-2\psi(1)
\nonumber\\[1mm]
&=&\psi(1-2s-z_{kk+1}\partial_k)+\psi(2\bar s+\bar z_{kk+1}\bar\partial_k)-2\psi(1)
\nonumber\\[1mm]
&=&\psi(2s+z_{kk+1}\partial_k)+\psi(1-2\bar s-\bar z_{kk+1}\bar\partial_k)-2\psi(1)\,.
\end{eqnarray}
The  Hamiltonian $\widetilde{\mathcal{H}}_N$ is a self-adjoint operator, it commutes with the operators $D_N(u),\, \bar D_N(u)$
 as well with its twin,
$[\mathcal{H}_N,\widetilde{\mathcal{H}}_N]=0$.

The sum of the Hamiltonians can be written in the following form~\footnote{Deriving this representation we have used the
identity
similar those given in~\cite{Lipatov:1993qn,Lipatov:1998as}
$$2\ln[z]+[z]^{1-2s}\,\ln[i\partial]\, [z]^{2s-1}=[z]^{-2s}\,\ln[-iz^2\partial]\, [z]^{2s}=\ln[-i(z^2\partial+2sz)].$$}
\begin{equation}\label{H+HN}
H_N=\mathcal{H}_N+\widetilde{\mathcal{H}}_N=\ln[-i(z^2_1\partial_1+2sz_1)]+
\sum_{k=1}^{N-1} H_{kk+1}+\ln[i\partial_N]-2\psi(1)\,.
\end{equation}
The pair-wise Hamiltonians
\begin{eqnarray}
H_{kk+1}&=&\mathcal{H}_{kk+1}+\widetilde{\mathcal{H}}_{kk+1}
\nonumber \\
&=&
2\ln[z_{kk+1}]
+[z_{kk+1}]^{1-2s}\Big(\ln [i\partial_k]+\ln[i\partial_{k+1}]\Big)[z_{kk+1}]^{2s-1}-4\psi(1)
\end{eqnarray}
are $SL(2,\mathbb{C})$ invariant operators, $[S^{(0,\pm)}_{kk+1}, H_{kk+1}]=0$. They can be written in terms of the
operators of the conformal spins ${J}_{kk+1}$ and $\bar{{J}}_{kk+1}$ which are customary defined as follows
\begin{eqnarray}
{J}_{kk+1}({J}_{kk+1}-1)&=&S^{(+)}_{kk+1}S^{(-)}_{kk+1}+S^{(0)}_{kk+1}(S^{(0)}_{kk+1}-1)\,,
\nonumber\\
\bar{{J}}_{kk+1}(\bar{{J}}_{kk+1}-1)&=&\bar S^{(+)}_{kk+1}\bar S^{(-)}_{kk+1}+
\bar S^{(0)}_{kk+1}(\bar S^{(0)}_{kk+1}-1)\,.
\end{eqnarray}
The  Hamiltonian $H_{kk+1}$ as a function of the conformal spins ${J}_{kk+1}$, $\bar{{J}}_{kk+1}$ takes the standard form
\begin{equation}
H_{kk+1}=\psi\left({J}_{kk+1}\right)+\psi\left(1-{J}_{kk+1}\right)+\psi\left(\bar{{J}}_{kk+1}\right)
+\psi\left(1-\bar{{J}}_{kk+1}\right)-4\psi(1).
\end{equation}
For $s=0,\bar s=1$ the Hamiltonian~(\ref{H+HN}) coincides with the Hamiltonian obtained in~Ref.~\cite{Lipatov:2009nt}
which determines  the contribution of $N$-reggeized t-channel gluons to the scattering amplitudes in $N=4$ SUSY
(see Refs.~\cite{Bartels:2008sc,Bartels:2011nz,Lipatov:2009nt} for further details).
\vskip 5mm

The Hamiltonians, $\mathcal{H}_N$ and $ \widetilde{\mathcal{H}}_N$ belong to the commutative $D$-family. The
corresponding
eigenfunctions were constructed in Sect.~\ref{sect:iterative}, Eq.~(\ref{PsiD}). The eigenvalues of
$\mathcal{H}_N$ and $ \widetilde{\mathcal{H}}_N$ can be easily obtained from Eqs.~(\ref{QDexp}) and (\ref{QD-eigenvalue})
\begin{equation}
\mathcal{H}_N \,\Psi_{D}(\boldsymbol{x}|\boldsymbol{z})=E_N^{\boldsymbol{s}}(\boldsymbol{x})\,
\Psi_{D}(\boldsymbol{x}|\boldsymbol{z})\,,
\quad
\widetilde{\mathcal{H}}_N \,\Psi_{D}(\boldsymbol{x}|\boldsymbol{z})=
E_N^{1-\boldsymbol{s}}(\boldsymbol{x})\,\Psi_{D}(\boldsymbol{x}|\boldsymbol{z})\,,
\end{equation}
where
\begin{eqnarray}
E_N^{\boldsymbol{s}}(\boldsymbol{x})&=\sum_{k=1}^N \Big(\psi(1-s+ix_k)+\psi(\bar s-i\bar x_k)-2\psi(1)\Big)\,,
\nonumber\\
E_N^{1-\boldsymbol{s}}(\boldsymbol{x})&=\sum_{k=1}^N \Big(\psi(s+ix_k)+\psi(1-\bar s-i\bar x_k)-2\psi(1)\Big)\,.
\end{eqnarray}
Taking into account Eq.~(\ref{x-n-nu}) one gets
\begin{eqnarray}
E_N^{\boldsymbol{s}}(\boldsymbol{x})&=2\sum_{k=1}^N\mathrm{Re}
\left(\psi\left(\frac12+\frac{n_k-n_s}2+i(\nu_k-\nu_s)\right)-\psi(1)\right)\,,
\nonumber\\[2mm]
E_N^{1-\boldsymbol{s}}(\boldsymbol{x})&=
2\sum_{k=1}^N\mathrm{Re}\left(\psi\left(\frac12+\frac{n_k+n_s}2+i(\nu_k+\nu_s)\right)-\psi(1)\right)\,.
\end{eqnarray}
For $n_s=-1$, $\nu_s=0$ ($s=0,\bar s=1$) it agrees with the results of Ref.~\cite{Bartels:2008sc,Lipatov:2009nt}.
\section{Summary}\label{sect:summary}
We have developed an iterative method for the construction of eigenfunctions of the elements of the monodromy matrix
for the $SL(2,\mathbb{C})$ spin chains.  The whole construction relies heavily upon the properties of the operators which
factorize the $\mathcal{R}$-operator. The eigenfunctions are represented as the product of  operators that map the
functions of $k-$variables to the functions of $k+1$ variables. The integral kernels of these operators can be
represented in the form of two-dimensional Feynman diagrams. Using the diagrammatic technique we have calculated the scalar
products of the corresponding eigenfunctions and determined the so-called Sklyanin's measure.

We  have paid a special attention to the eigenfunctions of the $D_N$ operator.
These eigenfunctions describe bound states of the regeized gluons  corresponding to the Regge cut contributions  to the
scattering amplitudes in $N=4$  SUSY.
We constructed set of Baxter operators (commutative families) which commute with the corresponding elements of the
monodromy matrix and studied their properties. It was shown that the Baxter operators
satisfy the first-order difference equation in the
spectral parameters. The eigenvalues of the Baxter operators were obtained in the explicit form.
Expanding the Baxter operator at the special  point we obtained two
self-adjoint Hamiltonians that belong to the commutative $D-$family. For the special choice of the conformal spins
($SL(2,\mathbb{C})$ representations) the sum of these Hamiltonians coincides with the Hamiltonian governing evolution of
reggeized gluons.

More generally our approach is based on the properties of factorizing operators and has to be applicable for
generic  models with
a factorizable $\mathcal{R}$ matrix.

\ack

This work  was supported
by RFBR grants 12-
02-91052, 13-01-12405, 14-01-00341 (S.D.) and
by the DFG grant BR2021/5-2 (A.M.).

\appendix

\section*{Appendices}
\addcontentsline{toc}{section}{Appendices}

\renewcommand{\theequation}{\Alph{section}.\arabic{equation}}
\renewcommand{\thetable}{\Alph{table}}
\setcounter{section}{0}
\setcounter{table}{0}
\section{Diagram technique}\label{Diagram}
\begin{figure}[t]
\centerline{\includegraphics[width=0.7\linewidth]{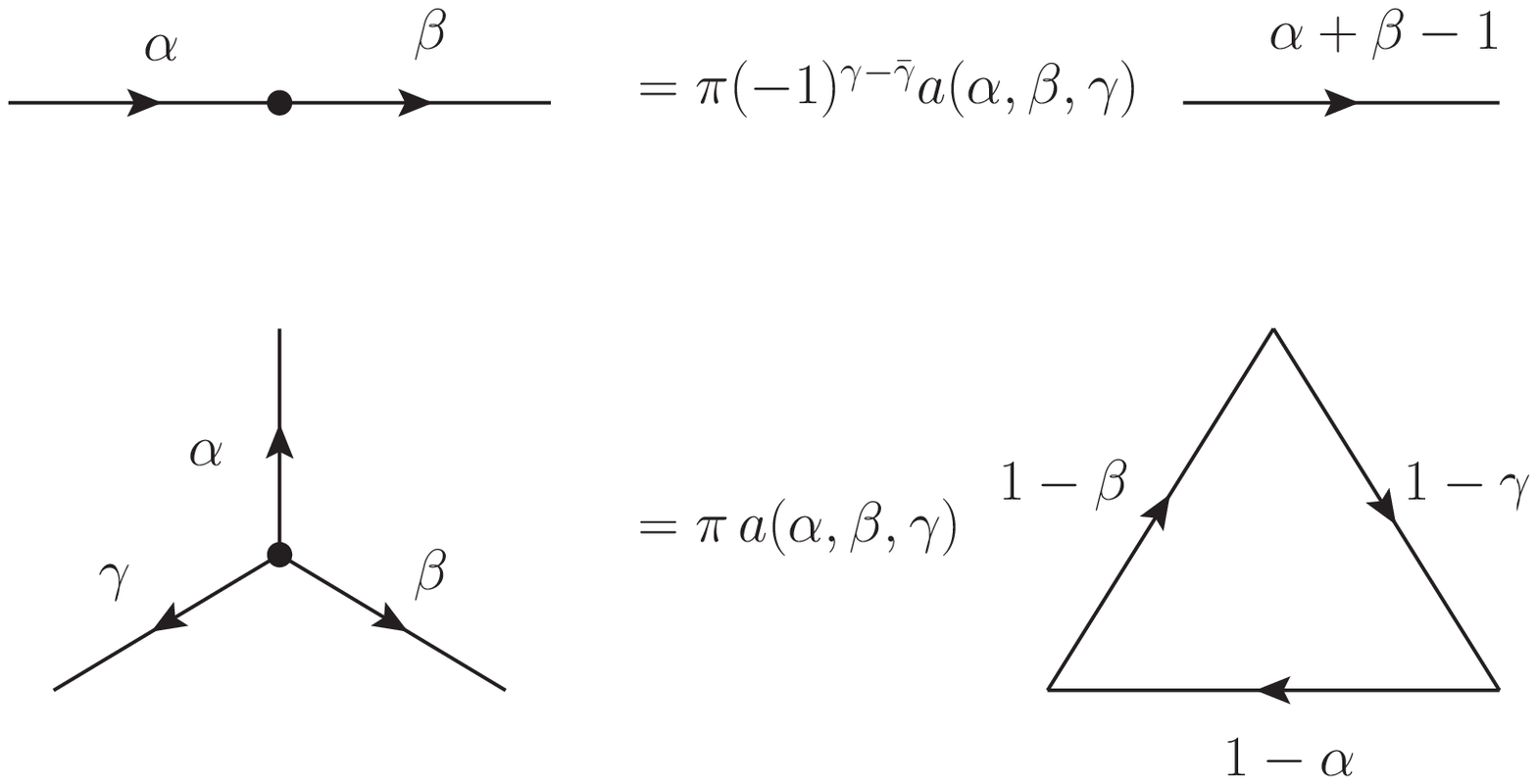}}
\caption{The chain and star--\,triangle relations, $\alpha+\beta+\gamma=2$.}
\label{Chain+Star}
\end{figure}

In this Appendix we present the basic elements of the diagram technique which was used throughout the paper.
The functions and  kernels of operators  considered in the main body of the paper are represented in the form of
two-dimensional Feynman diagrams. The propagator which is shown by the arrow directed from $w$ to $z$ and
index α attached to it as shown in Fig.~\ref{Propagator} is given by the following expression
\begin{equation}
\frac{1}{[z-w]^\alpha}\equiv\frac{1}{(z-w)^\alpha (\bar z-\bar w)^{\bar\alpha}}=
\frac{(\bar z-\bar w)^{\alpha-\bar\alpha}}{|z-w|^{2\alpha}}=\frac{(-1)^{\alpha-\bar\alpha}}{[w-z]^{\alpha}}\,,
\end{equation}
where $\alpha-\bar\alpha=n_\alpha$ is integer. Making the Fourier transformation we define
the propagator in the momentum representation
\begin{equation}\label{Fourier}
\int d^2 z \frac{e^{i(pz+\bar p\bar z)}}{[z]^\alpha}=\pi\, i^{\alpha-\bar\alpha}\, a(\alpha)\,\frac{1}{[p]^{1-\alpha}}\,.
\end{equation}
Here   the notation $a(\alpha)$ is introduced for the function
\begin{equation}
a(\alpha)=\frac{\Gamma(1-\bar\alpha)}{\Gamma(\alpha)}\,, \quad a(\bar\alpha)=\frac{\Gamma(1-\alpha)}{\Gamma(\bar\alpha)}\,,
\quad a(\alpha,\beta,\gamma,\ldots)=a(\alpha)a(\beta) a(\gamma)\ldots
\end{equation}
It has the following properties
\begin{equation*}\fl
a(\alpha) a(1-\bar\alpha)=1\,, \quad \frac{a(1+\alpha)}{a(\alpha)}=-\frac{1}{\alpha\bar\alpha}\,, \quad
a(\alpha)a(1-\alpha)=(-1)^{\alpha-\bar\alpha}\,,\quad  a(\alpha)=(-1)^{\alpha-\bar\alpha} a(\bar\alpha)\,.
\end{equation*}
Making use of Eq.~(\ref{Fourier}) it is easy to derive
the following useful representation for  the fractional derivative $[i\partial]^\alpha$,
\begin{equation}\label{FD}
\int d^2w \frac{1}{[z-w]^\alpha} f(w)=\pi(-i)^{\alpha-\bar\alpha}\, a(\alpha)\, [i\partial]^{\alpha-1}\, f(z).
\end{equation}

The evaluation of  Feynman diagrams is based on their transformation with the help of the certain "integration rules"

\begin{itemize}
\item Chain  relation:
\begin{equation}\label{Chain}\fl
\int d^2 w\frac{1}{[z_1-w]^\alpha [w-z_2]^{\beta}}=
(-1)^{\gamma-\bar\gamma}a(\alpha,\beta,\gamma)
\frac{1}{[z_1-z_2]^{\alpha+\beta-1}}\,,
\end{equation}
where $\gamma=2-\alpha-\beta,\ \bar\gamma=2-\bar\alpha-\bar\beta$.
\item Star--\,triangle relation:
\begin{equation}\label{Star}\fl
\int d^2w\frac{1}{[z_1-w]^\alpha[z_2-w]^\beta [z_3-w]^\gamma}=
\frac{\pi a(\alpha,\beta,\gamma)}{[z_2-z_1]^{1-\gamma}[z_1-z_3]^{1-\beta}[z_3-z_2]^{1-\alpha}}\,,
\end{equation}
where $\alpha+\beta+\gamma=2$ and $\bar\alpha+\bar\beta+\bar\gamma=2$. In an operator form star-triangle relation
reads~\cite{Isaev:2003tk}
\begin{equation}\label{STOP}
[z]^\alpha\, [i\partial]^{\alpha+\beta}\, [z]^\beta=[i\partial]^{\beta}\,[z]^{\alpha+\beta}\, [i\partial]^{\alpha}\,.
\end{equation}

\item Cross relation:
\begin{eqnarray}\label{Cross}\fl
\frac{1}{[z_1-z_2]^{\alpha'-\alpha}}\int d^2w
\frac{a(\alpha',\bar\beta')}{[w-z_1]^\alpha[w-z_2]^{1-\alpha'}
[w-z_3]^\beta [w-z_4]^{1-\beta'}}=
\nonumber\\
=\frac{1}{[z_3-z_4]^{\beta'-\beta}}
\int d^2 w
\frac{a(\alpha,\bar\beta)}{[w-z_1]^{\alpha'}[w-z_2]^{1-\alpha}
[w-z_3]^{\beta'} [w-z_4]^{1-\beta}}\,,
\end{eqnarray}
where $\alpha+\beta=\alpha'+\beta'$.
\end{itemize}
\begin{figure}[t]
\centerline{\includegraphics[width=0.9\linewidth]{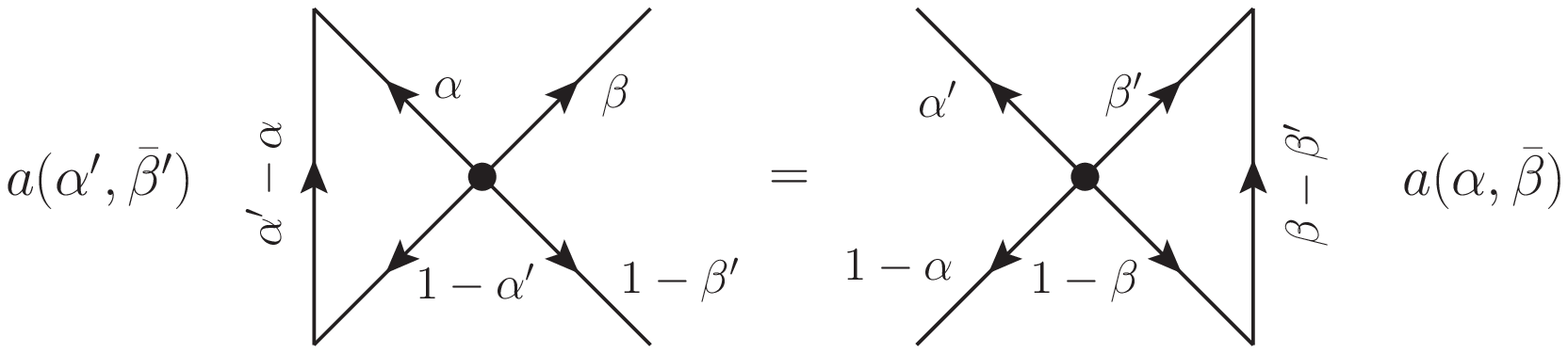}}
\caption{The cross relation, $\alpha+\beta=\alpha'+\beta'$.}
\label{fig:Cross}
\end{figure}
These relations are shown in diagrammatic form in Figs.~\ref{Chain+Star}, \ref{fig:Cross}.

Finally, we give  two  representations for   the  $\delta$ function.
The first one
\begin{equation}\label{delta1}
\delta^2(z)=\lim_{\epsilon\to 0}\frac{a(i\epsilon)}{\pi} \frac1{[z]^{1-i\epsilon}}=
\lim_{\epsilon\to 0}\frac{i\epsilon}{\pi}\frac1{[z]^{1-i\epsilon}}
\end{equation}
follows directly from Eq.~(\ref{Fourier}) and  the second relation
\begin{equation}\label{Delta2}
\int d^2 w\frac{1}{[z_1-w]^{2-\alpha}[w-z_2]^\alpha}=\pi^2\, a(\alpha,2-\alpha)\, \delta^2(z_1-z_2)\,
\end{equation}
results from the chain relation~(\ref{Chain}) and (\ref{delta1}).

\section{Proof of  commutativity}\label{App:B}
The proof of the commutativity of the ``transfer matrices'' $\mathcal{T}_{\boldsymbol{s}_0}(\boldsymbol{u})$, Eq.~(\ref{Om2})
is based on the Yang-Baxter relation
\begin{equation}\label{YBE}
\mathcal{R}_{\boldsymbol{s}_0\boldsymbol{s}_{0'}}(\boldsymbol{u}-\boldsymbol{v})
\mathbb{T}_{\boldsymbol{s}_0}(\boldsymbol{u})\mathbb{T}_{\boldsymbol{s}_{0'}}(\boldsymbol{v})=
\mathbb{T}_{\boldsymbol{s}_{0'}}(\boldsymbol{v})\mathbb{T}_{\boldsymbol{s}_0}(\boldsymbol{u})
\mathcal{R}_{\boldsymbol{s}_0\boldsymbol{s}_{0'}}(\boldsymbol{u}-\boldsymbol{v})
\end{equation}
and two special identities for the kernel of the operator
$\mathcal{R}_{\boldsymbol{s}_0\boldsymbol{s}_{0'}}(\boldsymbol{u}-\boldsymbol{v})$:
%
\begin{eqnarray}\label{2Ia}
&&\int d^2 w_1\, d^2w_2\,\mathcal{R}_{\boldsymbol{s}_0\boldsymbol{s}_{0'}}(\boldsymbol{u})(z_1,z_2|w_1,w_2)=
C(u,s_0,s_{0'})\,,
\\ \label{2Ib}
&&
\lim_{z_1,z_2\to z}\mathcal{R}_{\boldsymbol{s}_0\boldsymbol{s}_{0'}}(\boldsymbol{u})(z_1,z_2|w_1,w_2)=
C(u,s_0,s_{0'})\,\delta^2(z-w_1)\,\delta^2(z-w_2)\,,
\end{eqnarray}
%
where $C(u,s_0,s_0')$ is some coefficient. The kernels on the l.h.s and r.h.s. of Eq.~(\ref{YBE}) depend on the variables $(z_k, w_k)$,
$k=1,\ldots,N$
in the quantum space and the variables $z_0, z_{0'}, w_0,w_{0'}$ associated with the two auxiliary spaces.
Sending $z_0,z_{0'}\to 0$ and integrating over $w_0,w_{0'}$ in both parts of Eq.~(\ref{YBE}) with the help of Eqs.~(\ref{2Ia}), (\ref{2Ib})
 one immediately gets
Eq.~(\ref{Om2}).

The identities~(\ref{2Ia}), (\ref{2Ib}) follow from the analogous identities for the factorized operators
 $\mathcal{R}_{12}^{(k=1,2)}$, see Eq.~(\ref{Rint}):
%
\begin{eqnarray}\label{2Ra}
&&
\int d^2 w_1\, d^2w_2\,\mathcal{R}_{12}^{(k)}(u,v)(z_1,z_2|w_1,w_2) =
A^{(k)}(u,v)\,,
\\
\label{2Rb}
&&
\lim_{z_1\to z, z_2\to z}\mathcal{R}_{12}^{(k)}(u,v)(z_1,z_2|w_1,w_2)  = A^{(k)}(u)\,\delta^2(z-w_1)\,\delta^2(z-w_2)\,,
\end{eqnarray}
where $A^{(2)}(u,v)=A^{(1)}(v,u)$ and
\begin{equation}
A^{(1)}(u,v)=\pi\,(-1)^{i(v-\bar v)}\,a(iv,1-iu,1+iu-iv).
\end{equation}
To derive Eqs.~(\ref{2Ia}) or (\ref{2Ra}) it is sufficient to use the chain relation~(\ref{Chain}).
In order to obtain~(\ref{2Ra}) one has to represent the kernel in the form of the star diagram using
the star-triangle relation~(\ref{Star}) then send $z_1\to z_2$ and use the chain integration rule~(\ref{Delta2}).
Eq.~(\ref{2Ib}) follows from Eqs.~(\ref{2Rb}) and (\ref{Rmatrix}).

\end{document}